\documentclass[a4paper,fleqn]{cas-dc}
\usepackage{bbold}
\DeclareUnicodeCharacter{FF08}{(}

\usepackage{amsmath}
\usepackage{ragged2e}
\usepackage{amssymb}
\usepackage{caption} 
\usepackage{booktabs}
\usepackage{array, caption, threeparttable}
\usepackage[font=small,labelfont=bf,labelsep=none]{caption}
\usepackage[numbers]{natbib}
\usepackage{bbold} 
\newcommand{\myfont}{\fontfamily{ptm}\selectfont} 

\def\tsc#1{\csdef{#1}{\textsc{\lowercase{#1}}\xspace}}
\tsc{WGM}
\tsc{QE}
\usepackage{soul}
\soulregister\cite7 
\soulregister\citep7 
\soulregister\citet7 
\soulregister\ref7 
\soulregister\pageref7 
\soulregister\cal7
\soulregister\rm7
\usepackage{diagbox}  

\begin{document}
\let\WriteBookmarks\relax
\def\floatpagepagefraction{1}
\def\textpagefraction{.001}
\let\printorcid\relax 


\shorttitle{\fontsize{8.5}{12}\selectfont Intermediate Domain-guided Adaptation 
for Unsupervised Chorioallantoic Membrane Vessel Segmentation
}
\shortauthors{Song \textit{et al}.}

\title[mode=title]{Intermediate Domain-guided Adaptation 
for Unsupervised Chorioallantoic Membrane Vessel Segmentation}

\author[1, 2]{Pengwu Song}
\cormark[1]
\ead{pwsong@mail.ustc.edu.cn}

\author[1, 2]{Zhiping Wang}
\cormark[1] 
\ead{wzp962@mail.ustc.edu.cn}

\author[3]{Peng Yao}
\cormark[2]
\ead{yaopeng@ustc.edu.cn}

\author[1, 2]{Liang Xu}
\ead{xul66@mail.ustc.edu.cn}

\author[1, 2]{Shuwei Shen}
\cormark[2]
\ead{swshen@ustc.edu.cn}

\author[4]{Pengfei Shao}
\ead{spf@ustc.edu.cn}

\author[1, 2]{Mingzhai Sun}
\ead{mingzhai@ustc.edu.cn}

\author[1, 2, 4]{Ronald X.Xu}
\cormark[2]
\ead{xux@ustc.edu.cn}

\address[1]{School of Biomedical Engineering, Division of Life Sciences and Medicine, University of Science and Technology of China, Hefei, Anhui,  \\230026, P.R.China}
\address[2]{Suzhou Institute for Advanced Research, University of Science and Technology of China, Suzhou, Jiangsu, 215123, P.R.China}
\address[3]{School of Microelectronics, University of Science and Technology of China, Hefei 230026, P.R.China}
\address[4]{Department of Precision Machinery and Precision Instrument, University of Science and Technology of China, Hefei 230026, P.R.China}

\cortext[1]{These authors contributed equally to this work.}
\cortext[2]{Corresponding author}

\begin{abstract}
The chorioallantoic membrane (CAM) model is a widely used in vivo platform for studying angiogenesis, especially in relation to tumor growth, drug delivery, and vascular biology.  Since the topology and morphology of developing blood vessels is a key evaluation metric, accurate vessel segmentation is essential for quantitative analysis of angiogenesis. However, manual segmentation is extremely time-consuming, labor-intensive, and prone to inconsistency due to its subjective nature. Moreover, research on CAM vessel segmentation algorithms remains limited, and the lack of public datasets contributes to poor prediction performance. To address these challenges, we propose an innovative Intermediate Domain-guided Adaptation (IDA) method, which utilizes the similarity between CAM images and retinal images, along with existing public retinal datasets, to perform unsupervised training on CAM images. Specifically, we introduce a Multi-Resolution Asymmetric Translation (MRAT) strategy to generate intermediate images to promote image-level interaction. Then, an Intermediate Domain-guided Contrastive Learning (IDCL) module is developed to disentangle cross-domain feature representations. This method overcomes the limitations of existing unsupervised domain adaptation (UDA) approaches, which primarily concentrate on directly source-target alignment while neglecting intermediate domain information. Notably, we create the first CAM dataset to validate the proposed algorithm. Extensive experiments on this dataset show that our method outperforms compared approaches. Moreover, it achieves superior performance in UDA tasks across retinal datasets, highlighting its strong generalization capability. The CAM dataset and source codes are available at https://github.com/Light-47/IDA.
\end{abstract}


\begin{highlights}

\item We propose a novel IDA method aimed at addressing challenges such as limited images and lack of labels in the CAM vessel segmentation task. To our knowledge, this is the first work to leverage labeled retinal images and unlabeled CAM images to assist in training a high-performance CAM vessel segmentation model.
\item We are the first to generate intermediate images at different resolutions in the unsupervised vessel segmentation domain. This enables image-level cross-domain interaction, allowing the model to balance both the overall and detailed vascular structures effectively.
\item To enhance feature discriminability, we design an IDCL module. This approach mitigates inter-domain discrepancies, facilitating effective feature-level adaptation without the need to explicitly minimize domain distances in the feature space.
\item To advance research on CAM vessel segmentation, we construct the first publicly available CAM dataset. Our method achieves state-of-the-art performance on the DRIVE → CAM\_DB task. Furthermore, it demonstrates superior generalization ability on the DRIVE → CHASEDB1 and DRIVE → STARE tasks.

\end{highlights}


\maketitle

\section{Introduction}
Angiogenesis is a strictly regulated process critical for embryonic development and plays a pivotal role in various pathophysiological conditions, such as wound healing, ischemic heart disease, and cancer \cite{nitzsche2022coalescent}. One of the most widely used assays for studying angiogenesis is the developing chick embryo and its chorioallantoic membrane (CAM) \cite{doukas2006automated}. Its popularity stems from several advantages, including the readily accessible experimental materials, short experimental duration, and the inherent immunodeficiency during the early culture stages \cite{guerra2021simulation, 2023Human}. Assessment of angiogenic activity in the developing CAM relies on key metrics such as vascular surface area, vessel length, and branching number. Therefore, the accuracy of vessel segmentation in CAM images is fundamental to assessing angiogenesis, offering critical insights into vascular development and related pathologies \cite{huang2015research}. For instance, tumors often induce vessel formation to support nutrient, waste transport, and gas exchange \cite{hanahan2011hallmarks}. In hypoxic tumor microenvironments, tumors stimulate excessive angiogenesis via pro-angiogenic factors \cite{liu2023angiogenic}. The CAM model, with its rich vascular network, is ideal for vascularization and angiogenesis studies. Combined with vessel segmentation, it quantifies tumor-induced vascular changes, offering key insights for anti-angiogenic drug evaluation. 

\begin{figure}
    \centering
    \includegraphics[width=8.3cm]{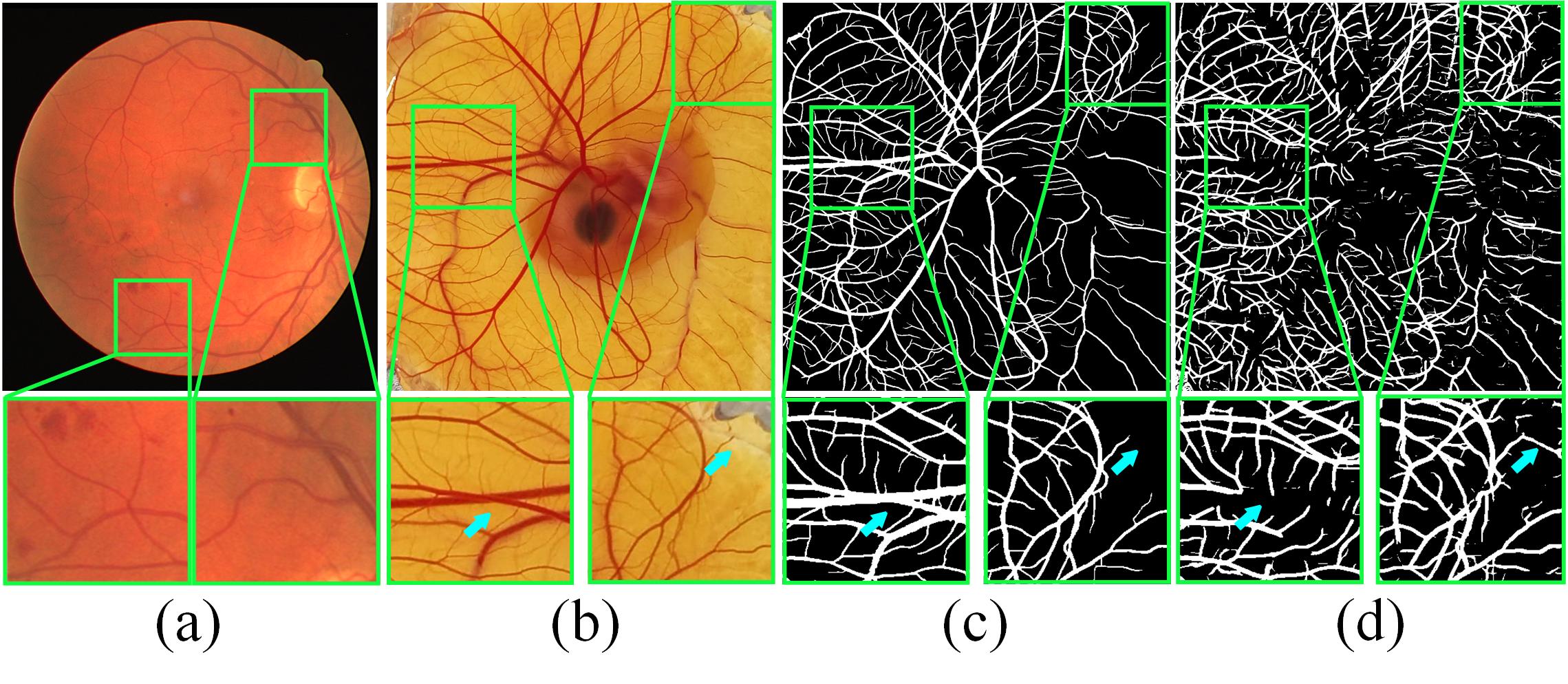}
    \caption{A comparison between retinal and CAM vessels, along with an example of performance degradation when applying a model trained on retinal images to CAM images. (a) Source retinal images. (b) Target CAM images. (c) Ground truth. (d) Predictions. Bottom left: under-segmentation. Bottom right: over-segmentation. The images below (a-d) show the corresponding enlarged views.}
    \label{figure1}
\end{figure}

In recent years, numerous studies \cite{zhou2021study, galdran2022state, tan2022retinal} have designed vessel segmentation algorithms based on supervised deep learning, achieving promising results. However, these methods often require extensive pixel-level annotations and mainly focus on retinal vessel segmentation, with little attention to CAM vessels. Similar to retinal images, CAM images exhibit dense vascular networks with complex branching and connectivity, making manual annotation highly time-consuming and labor-intensive. To our knowledge, no publicly available annotated CAM dataset has been released to date. If we apply the model trained on labeled source data (e.g., retinal images) directly to unlabeled target data like CAM images, it will lead to substantial performance degradation, as shown in Fig. \ref{figure1}, due to the notable discrepancies between the two image types.

Despite the significant image differences, can we leverage existing rich retinal datasets \cite{staal2004ridge, fraz2012ensemble, hoover2000locating} to assist in training a high-performance CAM vessel segmentation model$?$ To our knowledge, this is the first attempt to explore this approach. Typically, unsupervised domain adaptation (UDA) methods \cite{vu2019advent, feng2023unsupervised, liu2022margin, xie2023sepico, lu2024domain, kuang2023mscda, liu2024unsupervised} are used for model transfer between domains with distribution gaps. UDA reduces the distribution gap between source and target domains, improving model generalization from labeled source data to unlabeled target data. However, if the gap is too large, UDA may fail, leading to poor target domain performance. To demonstrate the feasibility of this approach, we analyze both CAM and retinal images, highlighting their topological similarities, particularly in the publicly available DRIVE retinal dataset \cite{staal2004ridge}. Based on these observations, we believe that employing UDA to aid in training a high-performance CAM segmentation model is a viable solution. To further validate this, we design a novel UDA-based framework for vessel segmentation, termed Intermediate Domain-guided Adaptation (IDA), which constructs intermediate samples between retinal and CAM vessel images to effectively mitigate domain shift. Extensive experiments confirm the effectiveness of our approach.

Early research on UDA segmentation tasks primarily focuses on reducing domain shifts through adversarial learning algorithms \cite{vu2019advent, feng2023unsupervised, liu2022margin, liu2024unsupervised, tsai2018learning, luo2019taking}. Recently, to minimize visual appearance differences across domains, researchers utilize image translation techniques, such as the Copy-Paste (CP), which synthesizes intermediate images by integrating the visual style of one domain with the semantic information of the other \cite{yang2020fda, lin2024unsupervised}. These simple CP methods not only effectively expand the dataset but also achieve the consistency regularization during model training. Moreover, when applied to UDA tasks, cross-domain CP can reduce domain shifts, further enhancing image-level adaptation. However, these methods are typically limited to intra-domain image generation or unidirectional translation between the source and target domains. This limitation results in insufficient utilization of intermediate domain information and suboptimal segmentation performance on the target domain \cite{bai2023bidirectional, ma2024constructing}. 

In addition, due to differences in image sizes between the source and target domains, as well as limitations in training resources and time, existing vessel segmentation methods \cite{zhou2021study, galdran2022state, tan2022retinal} typically adopt one of the following two preprocessing strategies. The first approach downsamples images to a uniform size, which struggles to capture fine structures, resulting in the inability of basic CP techniques to address the issue of blurry predictions for thin vessels. The second approach crops the images into smaller patches, focusing only on local regions. This carries the risk of losing global information and may disrupt the continuity of thicker vessels. To overcome these challenges, we propose a novel Multi-Resolution Asymmetric Translation (MRAT) strategy in our IDA framework. By combining the aforementioned preprocessing methods with two different CP operations (CutMix \cite{jiang2003adaptive} and ClassMix \cite{olsson2021classmix}), our approach effectively mitigates detail blurring, enhances vessel prediction continuity, while also achieving image-level adaptation.

To enhance feature expressiveness in the embedding space, contrastive learning has been widely applied into UDA \cite{feng2023unsupervised, liu2022margin, jiang2022prototypical}. These methods typically employ prototypes to establish unique latent representations for each class, thereby pulling features of the same class closer while pushing apart features of different classes. However, these methods mainly focus on single-domain representations, neglecting cross- domain feature distribution shifts. To address this problem, inspired by \cite{lu2024domain}, we introduce an innovative Intermediate Domain-guided Contrastive Learning (IDCL) module in our IDA framework, which leverages intermediate features generated by MRAT to enable both domain-level and feature-level adaptation.

To address the data scarcity, we construct a CAM vessel dataset named CAM\_DB
, comprising 335 images, including 40 with pixel-level annotations. Each image has a resolution of  $1024 \times 1024$. As the first publicly available CAM vessel dataset, it is expected to facilitate further research in this field. To identify a suitable source domain, we conduct preliminary transfer experiments using multiple retinal datasets (e.g., DRIVE \cite{staal2004ridge}, FIVES \cite{jin2022fives}) and find that DRIVE yields the best performance. Based on this, we adopt DRIVE as the source and carry out comprehensive evaluations against several representative UDA methods. Furthermore, to validate the generalization capability of our method, we also perform UDA experiments across different retinal datasets. The results show that our IDA achieves the best performance.

The main contributions are summarized as follows:

\textbf{1.} We propose a novel IDA method aimed at addressing challenges such as limited images and lack of labels in the CAM vessel segmentation task. To our knowledge, this is the first work to leverage labeled retinal images and unlabeled CAM images to assist in training a high-performance CAM vessel segmentation model.

\textbf{2.} Additionally, we are the first to generate intermediate images at different resolutions in the unsupervised vessel segmentation domain. This enables image-level cross-domain interaction, allowing the model to balance both the overall and detailed vascular structures effectively.

\textbf{3.} To enhance feature discriminability, we design an IDCL module. This approach mitigates inter-domain discrepancies, facilitating effective feature-level adaptation without the need to explicitly minimize domain distances in the feature space.

\textbf{4.} To advance research on CAM vessel segmentation, we construct the first publicly available CAM dataset. To identify the most suitable source domain for adaptation, we conduct extensive experiments using multiple retinal datasets on the CAM\_DB
 task. Based on these results, we select DRIVE as the optimal source domain. Our method then achieves state-of-the-art performance on the DRIVE → CAM\_{DB}
 task. Furthermore, it demonstrates superior generalization ability on the DRIVE → CHASEDB1 and DRIVE → STARE tasks.

\section{Related Works}
\subsection{Vessel Segmentation}
\textit{1) Traditional Methods:} Traditional vessel segmentation algorithms, such as matched filtering \cite{chaudhuri1989detection}, adaptive threshold \cite{yun2019cutmix}, and adaptive tracking \cite{liu1993recursive}, typically use morphological features or pixel intensity as reference indicators. As for CAM vessel, Doukas \textit{et al}. \cite{doukas2006automated} introduced an adaptive thresholding quantification algorithm; Bibiloni \textit{et al}. \cite{bibiloni2019vascular} proposed a fuzzy mathematical morphology-based segmentation method. Despite making significant progress, these methods often rely on predefined prior knowledge and require different parameter settings for each dataset.

\textit{2) Deep Learning Methods:} Unlike conventional methods, deep learning approaches can automatically generate segmentation reference indicators by combining features from different dimensions. For instance, Galdran \textit{et al}. \cite{galdran2022state} proposed a simple cascaded network architecture based on U-Net \cite{ronneberger2015u}, demonstrating its superiority through extensive experiments. Building on the network proposed in \cite{galdran2022state}, Tan \textit{et al}. \cite{tan2022retinal} developed a skeleton fitting module to preserve vessel morphology, and employed a contrastive loss to enhance the distinction between vessels and the background. In addition to pixel-level accuracy, recent supervised methods have emphasized preserving topological and structural properties of vessels. VSR-Net \cite{ye2025vsr} employed graph clustering to enhance connectivity, while Cheng \textit{et al}. \cite{cheng2021joint} jointly optimized topology and representation quality. Shi \textit{et al}. \cite{shi2022local} utilized local intensity order priors for curvilinear robustness, and Stucki \textit{et al}. \cite{stucki2023topologically} incorporated persistent homology to enforce topological faithfulness via barcode matching. However, these methods typically require a large amount of labeled data. Due to the domain shift, the trained models often struggle to generalize to other vessel datasets. As a result, researchers have increasingly turned to UDA-based methods for unlabeled vessel segmentation, which have demonstrated promising results \cite{liu2024unsupervised,lin2024unsupervised, feng2021unsupervised}.

\subsection{Unsupervised Domain Adaptation}
Unsupervised Domain Adaptation (UDA) aims to utilize source domains with abundant labeled data to address tasks on unlabeled target domains. Among the early efforts, adversarial learning has been widely adopted to encourage domain- invariant representations \cite{vu2019advent, feng2023unsupervised, liu2022margin, tsai2018learning, chen2019synergistic}. For instance, Chen \textit{et al}. \cite{chen2019synergistic} proposed a synergistic image-feature adaptation framework with adversarial objectives, while Liu \textit{et al}. \cite{liu2024unsupervised} developed a multi-level adversarial learning strategy specifically for cross-domain vessel segmentation. Additional advancements, such as entropy-based adversarial training \cite{vu2019advent, feng2023unsupervised} and output-structured discriminators \cite{liu2022margin, tsai2018learning}, further improve the consistency of segmentation predictions.
To further mitigate domain shifts, recent works have explored image-level \cite{lu2024domain, yang2020fda, na2021fixbi} and feature-level adaptation \cite{feng2023unsupervised, liu2022margin, xie2023sepico, cheng2023adpl} that explicitly minimize visual and representational discrepancies across domains. Image-level adaptation attempts to align inter-domain distributions based on certain visual representations, such as the image style. For example, Yang \textit{et al}. \cite{yang2020fda} introduced a Fourier-based transformation to synthesize intermediate domains, while Na \textit{et al}. \cite{na2021fixbi} proposed a fixed-ratio MixUp \cite{zhang2017mixup} to facilitate domain bridging through visual blending. Feature-level adaptation instead seeks to align distributions in the latent space. Cheng \textit{et al}. \cite{cheng2023adpl} proposed adaptive ClassMix \cite{olsson2021classmix} for bidirectional translation, and Feng \textit{et al}. \cite{feng2023unsupervised} introduced a semantic alignment module to reduce intra-class prototype distance.

Despite these advances, most UDA methods struggle to preserve fine-grained structures, particularly in cross-species biomedical scenarios, where both semantic consistency and topological fidelity are critical. To address these challenges, we propose a novel Multi-Resolution Asymmetric Translation (MRAT) strategy, aiming to enhance inter-domain interaction and improve the representation of fine image structures.

\subsection{Contrastive Learning}
Contrastive learning aims to bring same-class features closer and push different-class features apart in the embedding space, improving intra-class compactness and inter-class separation. It has been widely used in UDA \cite{liu2022margin, lu2024domain, kuang2023mscda, jiang2022prototypical}, proving its effectiveness. Liu \textit{et al}. \cite{liu2022margin} proposed a margin-preserving contrastive loss by introducing a deviation angle as a penalty for positive anchors to enhance inter-class differences. Kuang \textit{et al}. \cite{kuang2023mscda} extended the contrastive loss by combining pixel-to-pixel, pixel-to-prototype, and prototype-to-prototype contrasts to better utilize the underlying semantic information at different levels. However, existing methods focus on obtaining better discriminative representations within a single domain while neglecting inter-domain feature differences. Although Lu \textit{et al}. \cite{lu2024domain} proposed a domain-interactive contrastive learning module that incorporates cross-domain alignment into the feature space, directly using source and target domain prototypes cannot fully eliminate the domain shift. To address this, we propose an intermediate domain-guided contrastive learning module, which leverages intermediate features to guide cross-domain feature alignment.

\section{Method}
\subsection{Framework Overview}
\begin{figure*}
    \centering
    \includegraphics[width=17.35cm,height=7cm]{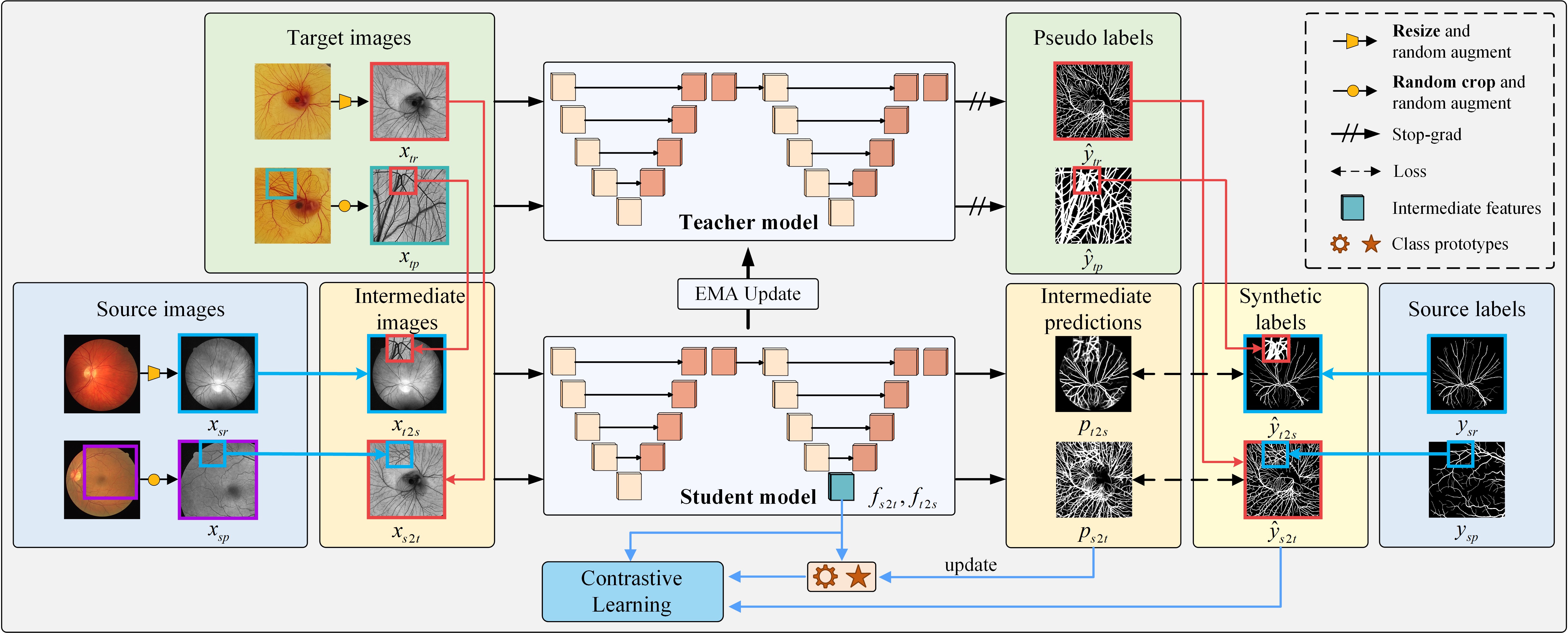}
    \caption{Framework of proposed method comprising two innovative components: (1) A multi-resolution asymmetric translation strategy that employs images of different resolutions for cross-domain transformation to generate intermediate images, facilitating image-level domain adaptation. (2) An intermediate domain-guided contrastive learning module, which adaptively updates prototypes using intermediate features and performs cross-domain contrastive learning with the intermediate prototypes to reduce domain gaps at the feature level.}
    \label{figure2}
    \vspace{-10pt}
\end{figure*}

The flowchart of our IDA method is illustrated in Fig. \ref{figure2}, which employs the teacher-student architecture \cite{tarvainen2017mean} as the basic framework. The student and teacher models share the same initialization, which is derived from pre-training on the source dataset. During training, the teacher model is updated by the student model using the Exponential Moving Average (EMA) strategy \cite{tarvainen2017mean}. We use $\phi_t$, $\phi_s$  to represent the parameters of the teacher model and the student model, respectively. The parameters of the teacher model are updated as follows:

\vspace{-15pt}
\begin{equation}
\label{equation1}
\varphi_t^{(k)} \leftarrow \lambda \varphi_t^{(k-1)} + (1-\lambda) \varphi_s^{(k)},
\end{equation}
\vspace{-15pt}

\noindent where $k$ denotes the number of iterations.  $\lambda$ is a momentum parameter \cite{tarvainen2017mean}. In this work, $\lambda$ is fixed to 0.99.

Considering accuracy and flexibility, we utilize cascaded U-Nets (W-Net) \cite{galdran2022state} as our backbone. The training set consists of a labeled source dataset $D^s = \{ x_s^i, y_s^j \}_{i=1}^N$ with $N$ retinal images and an unlabeled target dataset $D^t = \{x_t^i\}_{i=1}^M $ with $M$ CAM images, where $x_s^i$ and $x_t^i$  denote the $i$-th sample from the source and target domains, respectively, and $y_s^i$ represents the corresponding ground truth from the source domain. The training process can be summarized as follows: randomly select two source images  $\{ x_s^i, x_s^j \}$ and two target images  $\{ x_t^k, x_t^l \}$ from the training set. The four images are then processed through the multi-resolution asymmetric translation strategy to generate two intermediate images  $\{ x_{t2s}, x_{s2t} \}$, which are input to the student model. Simultaneously, the transformed target images are fed into the teacher model to generate pseudo-labels. These pseudo-labels are combined with source labels to produce synthetic labels  $\{ \hat{y}_{t2s}, \hat{y}_{s2t} \}$, which guide the training of the student model. During training, we introduce an intermediate domain-guided contrastive learning module. This module utilizes the feature maps $\{ f_{t2s}, f_{s2t} \}$ from the student model’s encoder and the predicted outputs $\{ p_{t2s}, p_{s2t} \}$  from the student model’s decoder to jointly update prototypes. Subsequently, the contrastive loss is calculated using the intermediate features to ensure that the network fully exploits intermediate domain information, achieving feature-level cross-domain alignment and reducing domain gaps.

\subsection{Multi-Resolution Asymmetric Translation}
Specifically, as shown in Fig. \ref{figure2}, for two randomly selected labeled source images $\{ x_s^i, x_s^j \}$ and two unlabeled target $\{ x_t^k, x_t^l \}$ in the training set, we resize $x_s^i$ and $x_t^k$ to a uniform size of  $W \times H \times D$, while randomly cropping $x_s^j$ and $x_t^l$ to the same size of $W \times H \times D$, where $W$, $H$, and $D$ represent height, width, and channel, respectively. Due to significant color differences between the source and target images (Fig. \ref{figure1}), which can affect model performance, we convert all images to grayscale to eliminate color bias, resulting in images of size $W \times H$. To further increase data diversity, we apply data augmentation techniques to these four images, such as random flipping and color jitter. After these transformations, the four images are represented as $x_{sr}$, $x_{sp}$, $x_{tr}$, and $x_{tp}$.

To align the distributions of the source and target domains, we generate intermediate images by randomly replacing a region of the target image with corresponding source image (or vice versa), while keeping the resolution and overall size of the replaced images unchanged. In detail, we first create a CutMix mask $M_R$ by randomly setting a region of size $m \times m\ (m < \min\{W, H\})$ within an all-zero area of size $W \times H$ to 1. Then, we multiply the mask $M_R$ with the ground truth $y_{sp}$ of the source patch $x_{sp}$ resulting in a foreground-class ClassMix mask $M_C$. The two intermediate images can then be generated using the following equations:

\vspace{-20pt}
\begin{equation}
\label{equation2}
x_{t2s} = x_{tp} \odot M_R + x_{sr} \odot (\mathbf{1} - M_R),
\end{equation}
\vspace{-20pt}

\vspace{-20pt}
\begin{equation}
\label{equation3}
x_{s2t} = x_{sp} \odot M_C + x_{tr} \odot (\mathbf{1} - M_C),
\end{equation}
\vspace{-20pt}

\noindent where $\mathbf{1} \in \{1\}^{W \times H}$, $\odot$ means element-wise multiplication. These two intermediate images are then fed into the student model for training. Supervisory signals are also generated in a similar manner, as shown in the following equation:

\vspace{-20pt}
\begin{equation}
\label{equation4}
\hat{y}_{t2s} = \hat{y}_{tp} \odot M_R + y_{sr} \odot (\mathbf{1} - M_R),
\end{equation}
\vspace{-20pt}

\vspace{-20pt}
\begin{equation}
\label{equation5}
\hat{y}_{s2t} = y_{sp} \odot M_C + \hat{y}_{tr} \odot (\mathbf{1} - M_C),
\end{equation}
\vspace{-20pt}

\noindent where $\hat{y}_{t2s}$, $\hat{y}_{s2t}$ denote the synthetic labels of the intermediate images, ${y}_{sr}$, ${y}_{sp}$ are the ground truths of ${x}_{sr}$, ${x}_{sp}$, and $\hat{{y}}_{tr}$, $\hat{{y}}_{tp}$ correspond to the pseudo-labels of ${x}_{tr}$, ${x}_{tp}$, which are obtained from the teacher model through an argmax operation followed by one-hot encoding.

Since the patches pasted in ${x}_{t2s}$ and ${x}_{s2t}$ are taken directly from ${x}_{tp}$ and ${x}_{sp}$ without downsampling the original images, thin vessel details are preserved. Meanwhile, the remaining regions are derived from ${x}_{sr}$ and ${x}_{tr}$, retaining the overall structural information of the source and target images. Through these unique CP operations, our approach can efficiently alleviate detail blurring and improve the continuity of vessel prediction. Additionally, it is worth noting that for ${x}_{t2s}$, since ground truths are unavailable, we employ a CutMix-like operation, directly pasting a patch from  ${x}_{sp}$. For ${x}_{s2t}$, we use a ClassMix-like operation, pasting only the foreground from the patch of  ${x}_{sp}$ onto ${x}_{tr}$. This overlay of the vessel foreground from both ${x}_{sp}$ and ${x}_{tr}$ creates an interlacing effect of vascular structures, enabling the model to better handle such complex scenarios during training. The two intermediate images obtained by these two different methods serve two purposes: first, they enrich the diversity of the input and alleviate issues such as label blurring at the rectangular boundaries introduced by CutMix and the high accuracy requirement for labels in ClassMix. Second, they help reduce domain shifts from different perspectives. By applying consistency regularization with pseudo-labels during training, this approach enhances the model performance and generalization ability.

\subsection{Intermediate Domain-guided Contrastive Learning}
\begin{figure}
    \centering
    \includegraphics[width=8.3cm]{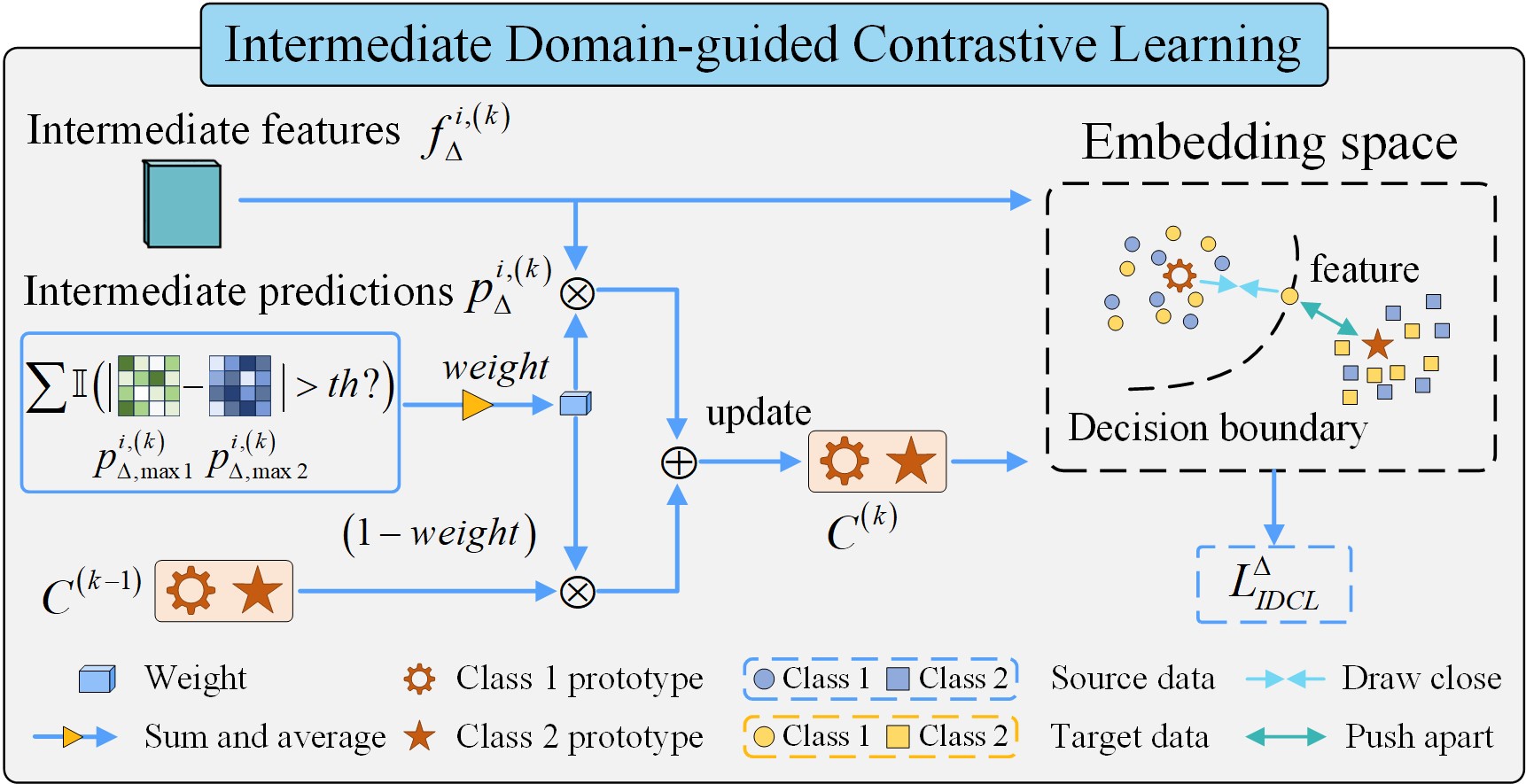}
    \caption{The diagram of the intermediate domain-guided contrastive learning module. First, the update weights at the $k$-th iteration are computed using the predictions $p^{i,(k)}_{\Delta}$ of the intermediate images. The prototypes $c^{(k)}$ is then updated by the weights and intermediate features $f^{i,(k)}_{\Delta}$. Finally, calculate the contrastive loss.}\label{figure3}
    \vspace{-10pt}
\end{figure}

The MRAT strategy described above generates unique intermediate images to facilitate image-level adaptation, while our IDCL method leverages prototype-based contrastive learning to further achieve feature-level adaptation. Unlike previous work \cite{liu2022margin, xie2023sepico, kuang2023mscda}, which relies on source and target features, our method uses intermediate features to update prototypes and perform contrastive learning. As illustrated in Fig. \ref{figure3}, this approach not only achieves intra-class compactness and inter-class separability within each domain but also reduces domain gaps, promoting better inter-domain interaction. Specifically, we first obtain the initial prototypes $C^{(0)} = \left\{ c_1^{(0)}, c_2^{(0)}, \dots, c_L^{(0)} \right\}$ by computing the class centers of the initial source pixel features from a pre-trained model on source images, where $L$ denotes the total number of classes ( $L=2$ for the vessel segmentation task). The prototypes are defined as follows:

\vspace{-17pt}
\begin{equation}
\label{equation6}
c_{r}^{(0)} = \frac{1}{|N_r|} \sum_{i=1}^{N} \sum_{j=1}^{h \times w} f_{s}^{i,(0)}[j] \cdot \mathbb{1}\!\left( Y_{s}^{i}[j] = r \right),
\end{equation}
\vspace{-12pt}

\noindent where $N$ represents the number of source samples.  $f_s^{i,(0)}$ denotes the feature map of the source image obtained through the pre-trained model, coming from the intermediate layer of the second U-Net model in the backbone (as shown in the “Intermediate features” of Fig. \ref{figure2}). $h$, $w$ are the height and width of the feature map, respectively.  $Y_{s}^{i}$ is obtained by downsampling the source label $y_{s}^{i}$ to a size of $h \times w$. $\mathbb{1}(\circ)$ is the indicator function, which equals 1 if $(\circ)$ is true. $|N_{r}|$ denotes the number of features belonging to the $r$-th class in the source domain, i.e. $|N_{r}| = \sum_{i=1}^{N} \sum_{j=1}^{h \times w} \mathbb{1}\!\left( Y_{s}^{i}[j] = r \right).$

To enhance the representational ability of prototypes and align the source and target domains at the feature level, we use the features generated from the intermediate images to update the prototypes:

\vspace{-17pt}
\begin{equation}
\label{equation7}
c_{r}^{(k')} \leftarrow (1 - w_{t2s}^{(k)}) c_{r}^{(k-1)} + w_{t2s}^{(k)} c_{r, t2s}^{(k)},
\end{equation}
\vspace{-12pt}

\vspace{-25pt}
\begin{equation}
\label{equation8}
c_{r}^{(k)} \leftarrow (1 - w_{s2t}^{(k')}) c_{r}^{(k)} + w_{s2t}^{(k)} c_{r, s2t}^{(k)},
\end{equation}
\vspace{-12pt}

\noindent where $k$ denotes the $k$-th iteration during training. $c_{r}^{(k')}$ denotes the intermediate variable for the $k$-th update.  $c_{r, t2s}^{(k)}$, $c_{r, s2t}^{(k)}$ represent the prototypes of the $r$-th class computed from intermediate source-like images and target-like images, while $w_{t2s}^{(k)}$, $w_{s2t}^{(k)}$  denote their respective update weights. These can be computed using the following formulas:

\vspace{-17pt}
{\small{
\begin{equation}
\label{equation9}
c_{r,\Delta}^{(k)} = \frac{1}{|B_{r,\Delta}|} \sum_{i=1}^{B_{\Delta}} \sum_{j=1}^{h \times w} f_{\Delta}^{i,(k)} [j] \cdot \mathbb{1} \left( \hat{Y}_{\Delta}^{i,(k)} [j] = r \right),
\end{equation}
}
}

\vspace{-25pt}
\small{
\begin{equation}
\label{equation10}
w_{\Delta}^{(k)} = \frac{1}{|B_{\Delta}|} \sum_{i=1}^{B_{\Delta}} \sum_{j=1}^{H \times W} \mathbb{1} \left( \left| P_{\Delta}^{i,(k)} [j]_{\max_1} - P_{\Delta}^{i,(k)} [j]_{\max_2} \right| > th_{\Delta} \right)
\end{equation}
}
\vspace{-12pt}

\noindent Where $ \Delta \in \{ t2s, s2t \}$.  $f_\Delta^{i, (k)}$ represents the feature map of the $i$-th intermediate image in each iteration. $\hat{Y}_{\Delta}^{i,(k)}$ is obtained by downsampling the synthesized labels $\hat{y}_{\Delta}^{i,(k)}$ to the size of  $h \times w$. $B_\Delta$ represents the number of intermediate images in an iteration. In this paper, $B_{t2s} = B_{s2t}$, both of which are half the batch size. $|B_{\Delta}|$ indicates the total number of pixels in the predicted probability map for the entire batch of intermediate images, where $|B_{\Delta}| = B_{\Delta} \times H \times W$. $\left| B_{r,\Delta} \right| = \sum_{i=1}^{B_\Delta} \sum_{j=1}^{h \times w} \mathbb{1} \left( \hat{Y}_\Delta^{i,(k)}[j] = r \right)$.  $p_{\Delta}^{i,(k)}$ represents the predicted probability map of the intermediate images, with the indices “$\max_1$” and “$\max2$” referring to the maximum and submaximum predicted probability values, respectively.  $th_{\Delta}$ is the threshold.

It can be observed that the update weight $w_{\Delta}^{(k)}$ is computed by considering only pixels with high prediction probability confidence. This approach aims to dynamically adjust $w_{\Delta}^{(k)}$  to mitigate the impact of unreliable pseudo-labels and features on the prototypes. When the model is initially transferred to the intermediate domain for training, both the reliability of features and pseudo-labels are low due to the domain shift. In this case,  $w_{\Delta}^{(k)}$ is relatively small, and the update of the prototypes is more conservative. As training progress, the reliability of features and pseudo-labels improves, leading to more high-confidence pixels. Consequently, as $w_{\Delta}^{(k)}$ increases, the prototypes can be updated more rapidly. By employing this dynamic prototype update method, we can obtain more expressive and generalized prototypes. Moreover, these intermediate prototypes are treated as the mean of the source and target domain features in the embedding space, which further guide subsequent contrastive learning to learn domain-invariant representations.

Given the prototype $c_{r}^{(k)}$ for class $r$ and intermediate pixel features  $f_\Delta^{i, k}[j]$, their cosine similarity are defined as follows: 

\vspace{-10pt}
\begin{equation}
\label{equation11}
\cos\left(\theta^{i,(k)}_{\Delta}[j,r]\right) = \frac{\left[c_r^{(k)}\right]^T f_{\Delta}^{i,(k)}[j]}{\left\|c_r^{(k)}\right\|_2 \left\|f_{\Delta}^{i,(k)}[j]\right\|_2},
\end{equation}
\vspace{-10pt}

\noindent where $\|\cdot\|_2$ denotes the $L_2$ regularization. By calculating the cosine similarity, we can then obtain the contrastive loss for the $i$-th intermediate image, defined as:

\vspace{-10pt}
\begin{equation}
\label{equation12}
\mathcal{L}_{\text{IDCL}}^{\Delta} = -\sum_{j=1}^{h\times w} \log \frac{\exp\left(\cos\left(\theta_{\Delta}^{i,(k)}[j,\hat{y}_j] + \delta\right)/\tau\right)}{\exp\left(\cos\left(\theta_{\Delta}^{i,(k)}[j,\hat{y}_j] + \delta\right)/\tau\right) + S(j)},
\end{equation}

\noindent where $\cos\left(\theta_{\Delta}^{i,(k)}[j,\hat{y}_j]\right)$ indicates the cosine similarity between the pixel feature and positive anchor (i.e., positive pair). $\hat{y}_j$ denotes the class of the $j$-th pixel. $S(j) = \sum_{r=1, r \neq \hat{y}_j}^{L} \exp\left(\cos\left(\theta_{\Delta}^{i,(k)}[j,r]\right)/\tau\right)$ is used for normalization. Following MPSCL \cite{liu2022margin}, $\delta$ represents the deviation angle penalty, which is used to preserve the margin of the positive prototype. The temperature $\tau$ is typically set to 1 by default. The goal of $\mathcal{L}_{\text{IDCL}}^{\Delta}$ is to minimize the distance between positive pairs and maximize the distance between negative pairs in the embedding space, as shown in the right part of Fig.~\ref{figure3}. Furthermore, the strategy of guiding contrastive learning with intermediate prototypes leverages intermediate information, ensuring more consistent learning across the source and target domains, thereby promoting feature-level domain adaptation.

\begin{figure*}
    \centering
    \includegraphics[width=12cm]{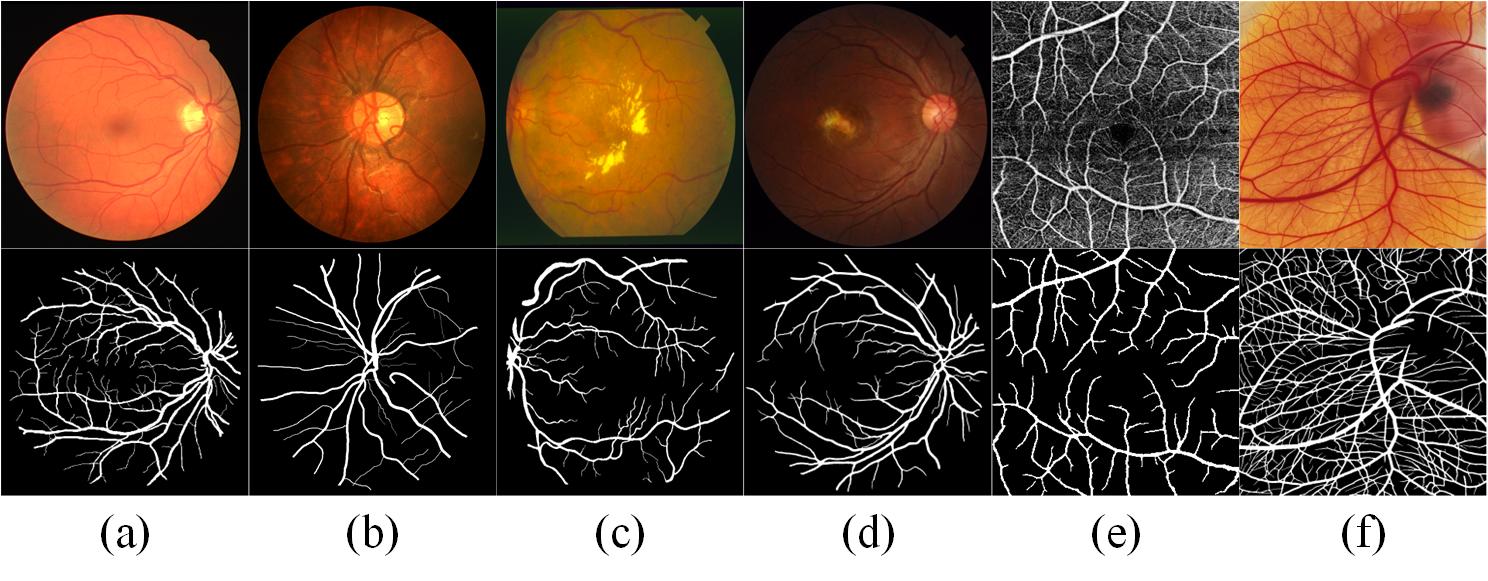}
    \caption{Visual comparison of different datasets. (a) DRIVE \cite{staal2004ridge}. (b) CHASEDB1 \cite{fraz2012ensemble}. (c) STARE \cite{hoover2000locating}. (d) FIVES \cite{jin2022fives}. (e) ROSSA \cite{ning2024accurate}. (f) CAM\_DB.}\label{figure4}
    \vspace{-10pt}
\end{figure*}

\subsection{Training Objective}

The loss for the intermediate samples during the training phase can be divided into three components: the supervised loss from the labeled source domain, the consistency loss from the unlabeled target domain, and the contrastive loss from the intermediate features. The supervised loss is further divided into the classification loss $\mathcal{L}_{cls}^s$ and segmentation loss $\mathcal{L}_{dice}^s$, which can be expressed as:

\vspace{-10pt}
\begin{equation}
\label{equation13}
    \mathcal{L}_{cls}^s = \mathcal{L}_{cls}\left(p_{t2s}, \hat{y}_{t2s}\right) \odot (\mathbf{1} - M_R) + \mathcal{L}_{cls}\left(p_{s2t}, \hat{y}_{s2t}\right) \odot M_C, \tag{13}
\end{equation}
\vspace{-10pt}

\vspace{-20pt}
\begin{equation}
\label{equation14}
    \mathcal{L}_{dice}^s = \mathcal{L}_{dice}\left(p_{t2s}, \hat{y}_{t2s}\right) \odot (\mathbf{1} - M_R) + \mathcal{L}_{dice}\left(p_{s2t}, \hat{y}_{s2t}\right) \odot M_C, \tag{14}
\end{equation}
\vspace{-10pt}

\noindent where $\mathcal{L}_{cls}$ represents the cross-entropy loss and $\mathcal{L}_{dice}$ denotes the DICE loss, which is usually applied in medical image segmentation tasks \cite{liu2022margin}. The consistency loss $\mathcal{L}_{con}$ uses the Mean Squared Error (MSE) loss, defined as follows:

\vspace{-10pt}
\begin{equation}
\label{equation15}
    \mathcal{L}_{con} = \mathrm{MSE}\left(p_{s2t}, p_{tr}\right) \odot (\mathbf{1} - M_C) + \mathrm{MSE}\left(p_{t2s}, p_{tp}\right) \odot M_R, \tag{15}
\end{equation}
\vspace{-10pt}

\noindent where $p_{tr}$, $p_{tp}$ represent the predicted probability maps of the target images $x_{tr}$ and $x_{tp}$ output by the teacher model.

In conclusion, the overall loss is summarized as follows:

\vspace{-10pt}
\begin{equation}
\label{equation16}
    \mathcal{L}_{all} = \mathcal{L}_{cls}^s + \mathcal{L}_{dice}^s + \beta_1 \mathcal{L}_{IDCL}^{t2s} + \beta_2 \mathcal{L}_{IDCL}^{s2t} + \gamma \mathcal{L}_{con}, \tag{16}
\end{equation}
\vspace{-10pt}

\noindent where $\beta_1$, $\beta_2$, $\gamma$ are hyperparameters used to balance the training. In this work, all are set to 1.

\section{Experiments}
\subsection{Experimental Setup}
\textbf{\textit{1) Datasets:}} In the experiments, we use five widely used retinal datasets ---DRIVE \cite{staal2004ridge}, CHASEDB1 \cite{fraz2012ensemble}, STARE \cite{hoover2000locating}, FIVES \cite{jin2022fives}, ROSSA \cite{ning2024accurate}---and a homemade CAM vessel dataset, CAM\_DB.

\textbf{DRIVE:} The DRIVE dataset is a publicly available dataset consisting of 40 color fundus images, accompanied by 40 corresponding segmentation labels. Each image has a resolution of $584 \times 565$. In our experiments, we uniformly use the DRIVE dataset as the source domain. We randomly select 36 images for the training set and the remaining 4 images for the validation set. 

\textbf{CHASEDB1:} The CHASEDB1 dataset consists of 28 color fundus images of both left and right eyes. Each image has a resolution of $999\times960$. In this study, we use 21 images for the training set, with the remaining images used for testing. 

\textbf{STARE:} The STARE dataset consists of 20 color fundus images, with a resolution of $700\times605$. We use 15 images for the training set, with the remaining 5 images used for testing.

\textbf{FIVES:} The FIVES dataset comprises 800 high-resolution $2048\times2048$ color fundus images, equally distributed across four categories: diabetic retinopathy (DR), age-related macular degeneration (AMD), glaucoma, and healthy controls (200 images each). In our experiments, 600 images are used for training, while the remaining 200 are reserved for testing.

\textbf{ROSSA:} The ROSSA dataset contains 918 labeled OCTA images, with a resolution of $320\times320$. We use 718 images for the training set, with the remaining 200 images used for testing.

\textbf{CAM\_DB:} After approximately 50 hours of incubation, we extract the embryos from the eggshells and place them in cups lined with plastic film. We then collect images of the chick embryos after 3 to 9 days of cultivation. To obtain vessel annotations, we first apply W-Net \cite{galdran2022state} for coarse segmentation, followed by manual refinement performed by two experts using ImageJ \cite{schneider2012nih}. The refined masks are subsequently validated by another expert. Due to the challenges in both image acquisition and annotation, we ultimately obtained 335 images, among which 40 are provided with pixel-level annotations. Each image has a resolution of $1024 \times 1024$. We use 295 unlabeled images as the training set and the remaining 40 labeled images as the test set.

As illustrated in Fig. \ref{figure4}, representative samples from different datasets exhibit varying degrees of domain shift with respect to CAM\_DB. The ROSSA dataset, based on OCTA imaging, presents the most significant modality gap. CHASEDB1, STARE, and FIVES include a wide range of pathological features, introducing greater variability and structural differences. Although DRIVE also contains diseased cases, its images generally exhibit clearer vessel structures and more uniform quality, rendering it visually more aligned with the CAM\_DB dataset. Table \ref{Table1} provides a quantitative assessment of the distributional differences between the source datasets and the target domain CAM\_DB. We employ Jensen–Shannon Divergence (JS) and Maximum Mean Discrepancy (MMD) to measure inter-dataset discrepancies, with features consistently extracted using a ResNet-50 \cite{he2016deep} pretrained model to ensure comparability and robustness. Despite the presence of substantial domain differences, DRIVE demonstrates relatively smaller discrepancies with CAM\_DB than the other four datasets, as evidenced by its lowest JS and MMD values, thereby aligning well with its visual consistency observed earlier.

\begin{table}
\centering
\caption{
Distributional similarity analysis of different source datasets with CAM\_DB.}
\myfont
\fontsize{2.2pt}{3pt}\selectfont  
\resizebox{\columnwidth}{!}{ 
\begin{tabular}{c !{\vrule width 0.15pt} c c}
\Xhline{0.3pt}
\rowcolor[HTML]{C0C0C0} 
Source & JS $\downarrow$ & MMD $\downarrow$ \\
\Xhline{0.3pt}
DRIVE \cite{staal2004ridge} & \textbf{0.0200} & \textbf{0.1810} \\
CHASEDB1 \cite{fraz2012ensemble} & 0.0241 & 0.2067 \\
STARE \cite{hoover2000locating} & 0.0291 & \underline{0.1842} \\
FIVES \cite{jin2022fives} & \underline{0.0204} & 0.1876 \\
ROSSA \cite{ning2024accurate} & 0.0260 & 0.3487 \\
\Xhline{0.3pt}
\end{tabular}}
\label{Table1}
\end{table}

\textbf{\textit{2) Evaluation Metrics: }}
In this paper, we evaluate algorithms performance from two complementary perspectives: pixel-level accuracy and topological structural correctness. 

Pixel-level accuracy: Following previous work \cite{zhou2021study,galdran2022state, liu2024unsupervised}, five commonly used evaluation measures are adopted: the Area under the Receiver Operating Characteristics Curve (AUC), Accuracy (ACC), Sensitivity (SE), Specificity (SP), and DICE score (DICE). Specifically, AUC measures overall discriminative ability across thresholds; ACC reflects overall pixel-wise correctness; SE emphasizes recall of vessel pixels, while SP assesses the suppression of false positives. DICE captures the overlap between prediction and ground truth. The combined use of these metrics enables a more comprehensive assessment of pixel-level segmentation performance.

Topological structural correctness: Segmentation performance in vessel analysis also critically depends on preserving vascular topology, especially in tasks like angiogenesis modeling. For this reason, we additionally report two structure-aware metrics: the Centerline Dice (clDICE) \cite{shit2021cldice}, which considers overlap along centerline structures and better captures continuity of fine vessels, and the Betti Matching Error (BM) \cite{stucki2023topologically}, which quantifies discrepancies in the topological characteristics (e.g., number of loops, connected components) between the predicted and ground truth masks.

These two categories of metrics reflect complementary aspects of segmentation quality: pixel-wise metrics assess local accuracy, whereas topology-aware metrics evaluate structural integrity. Their joint use provides a more balanced and reliable evaluation, particularly for vascular tasks where preserving anatomical connectivity is essential.

\begin{figure*}[htbp]
\centering
\setlength{\tabcolsep}{2pt}   

\begin{minipage}{\textwidth}
\centering
\captionsetup{type=table} 
\caption{Quantitative results of different methods on the DRIVE $\rightarrow$ CAM\_DB Task. The best result is marked as \textbf{bold}. ``$\downarrow$'' indicates lower is better.}
\label{Table2}
\myfont
\fontsize{10pt}{15pt}\selectfont
\resizebox{1\linewidth}{!}{
\begin{tabular}{c !{\vrule width 0.7pt} c c c c c c c}
\Xhline{1pt}
\rowcolor[HTML]{C0C0C0} 
Methods & AUC & ACC & SE & SP & DICE & clDICE & BM $\downarrow$ \\
\Xhline{1pt}
AdvEnt \cite{vu2019advent} & 0.9696$\pm$0.0034 & 0.9408$\pm$0.0052 & 0.8320$\pm$0.0094 & 0.9639$\pm$0.0058 & 0.8312$\pm$0.0128 & 0.8839$\pm$0.0040 & 0.5382$\pm$0.0093 \\
MPSCL \cite{liu2022margin} & 0.9701$\pm$0.0031 & 0.9415$\pm$0.0055 & 0.8258$\pm$0.0059 & 0.9601$\pm$0.0115 & 0.8319$\pm$0.0125 & 0.8799$\pm$0.0029 & 0.5377$\pm$0.0054 \\
SkelCon \cite{tan2022retinal} & \underline{0.9847$\pm$0.0016} & 0.9490$\pm$0.0037 & \underline{0.9316$\pm$0.0053} & 0.9527$\pm$0.0043 & \underline{0.8650$\pm$0.0087} & \underline{0.9200$\pm$0.0043} & \underline{0.5039$\pm$0.0194} \\
SEASA \cite{feng2023unsupervised} & 0.9700$\pm$0.0033 & 0.9416$\pm$0.0049 & 0.8292$\pm$0.0085 & \underline{0.9655$\pm$0.0054} & 0.8327$\pm$0.0122 & 0.8820$\pm$0.0037 & 0.5364$\pm$0.0164 \\
SePiCo \cite{xie2023sepico} & 0.9768$\pm$0.0017 & 0.9402$\pm$0.0035 & 0.8764$\pm$0.0176 & 0.9538$\pm$0.0061 & 0.8370$\pm$0.0080 & 0.8967$\pm$0.0054 & 0.5366$\pm$0.0093 \\
DCLPS \cite{lu2024domain} & 0.9833$\pm$0.0038 & \underline{0.9494$\pm$0.0060} & 0.9043$\pm$0.0275 & 0.9590$\pm$0.0055 & 0.8621$\pm$0.0166 & 0.9188$\pm$0.0043 & 0.5157$\pm$0.0170 \\
\Xhline{1pt}
\textbf{IDA (Ours)} & \textbf{0.9854$\pm$0.0010} & \textbf{0.9608$\pm$0.0021} & \textbf{0.9334$\pm$0.0027} & \textbf{0.9666$\pm$0.0025} & \textbf{0.8929$\pm$0.0052} & \textbf{0.9388$\pm$0.0015} & \textbf{0.4878$\pm$0.0059} \\
\Xhline{1pt}
\end{tabular}}
\end{minipage}

\vspace{1em} 

\begin{minipage}{\textwidth}
\centering
\captionsetup{type=table}
\caption{Quantitative results of different methods on the DRIVE $\rightarrow$ CHASEDB1 Task. The best result is marked as \textbf{bold}. ``$\downarrow$'' indicates lower is better.}
\label{Table3}
\myfont
\fontsize{10pt}{15pt}\selectfont
\resizebox{1\linewidth}{!}{
\begin{tabular}{c !{\vrule width 0.7pt} c c c c c c c}
\Xhline{1pt}
\rowcolor[HTML]{C0C0C0} 
Methods & AUC & ACC & SE & SP & DICE & clDICE & BM $\downarrow$ \\
\Xhline{1pt}
AdvEnt \cite{vu2019advent} & 0.9559$\pm$0.0013 & 0.9388$\pm$0.0007 & 0.7900$\pm$0.0114 & 0.9540$\pm$0.0019 & 0.7052$\pm$0.0011 & 0.7438$\pm$0.0034 & 0.3619$\pm$0.0197 \\
MPSCL \cite{liu2022margin} & 0.9601$\pm$0.0018 & 0.9417$\pm$0.0013 & 0.8087$\pm$0.0095 & 0.9552$\pm$0.0022 & 0.7198$\pm$0.0027 & 0.7722$\pm$0.0032 & 0.2700$\pm$0.0053 \\
SkelCon \cite{tan2022retinal} & 0.9695$\pm$0.0030 & 0.9461$\pm$0.0022 & 0.8443$\pm$0.0317 & 0.9565$\pm$0.0051 & 0.7436$\pm$0.0070 & 0.8006$\pm$0.0147 & 0.2440$\pm$0.0103 \\
SEASA \cite{feng2023unsupervised} & 0.9594$\pm$0.0018 & 0.9394$\pm$0.0019 & 0.8160$\pm$0.0124 & 0.9521$\pm$0.0034 & 0.7140$\pm$0.0038 & 0.7687$\pm$0.0046 & 0.2923$\pm$0.0066 \\
SePiCo \cite{xie2023sepico} & 0.9568$\pm$0.0010 & 0.9339$\pm$0.0009 & 0.8377$\pm$0.0102 & 0.9437$\pm$0.0018 & 0.7014$\pm$0.0025 & 0.7640$\pm$0.0030 & 0.3152$\pm$0.0124 \\
DCLPS \cite{lu2024domain} & \underline{0.9727$\pm$0.0012} & \underline{0.9477$\pm$0.0016} & \underline{0.8526$\pm$0.0069} & \underline{0.9574$\pm$0.0020} & \underline{0.7513$\pm$0.0054} & \underline{0.8094$\pm$0.0075} & \underline{0.2371$\pm$0.0119} \\
\Xhline{1pt}
\textbf{IDA (Ours)} & \textbf{0.9766$\pm$0.0005} & \textbf{0.9508$\pm$0.0007} & \textbf{0.8650$\pm$0.0076} & \textbf{0.9596$\pm$0.0013} & \textbf{0.7651$\pm$0.0026} & \textbf{0.8193$\pm$0.0041} & \textbf{0.2370$\pm$0.0073} \\
\Xhline{1pt}
\end{tabular}}
\end{minipage}

\vspace{1em} 

\begin{minipage}{\textwidth}
\centering
\captionsetup{type=table}
\caption{Quantitative results of different methods on the DRIVE $\rightarrow$ STARE task. The best result is marked as \textbf{bold}. ``$\downarrow$'' indicates lower is better.}
\label{Table4}
\myfont
\fontsize{10pt}{15pt}\selectfont
\resizebox{1\linewidth}{!}{
\begin{tabular}{c !{\vrule width 0.7pt} c c c c c c c}
\Xhline{1pt}
\rowcolor[HTML]{C0C0C0} 
Methods & AUC & ACC & SE & SP & DICE & clDICE & BM $\downarrow$ \\
\Xhline{1pt}
AdvEnt \cite{vu2019advent} & 0.9543$\pm$0.0049 & 0.9450$\pm$0.0014 & 0.7981$\pm$0.0153 & 0.9610$\pm$0.0031 & 0.7395$\pm$0.0026 & 0.8061$\pm$0.0035 & 0.2016$\pm$0.0145 \\
MPSCL \cite{liu2022margin} & 0.9535$\pm$0.0060 & 0.9480$\pm$0.0007 & 0.7814$\pm$0.0158 & 0.9661$\pm$0.0021 & 0.7461$\pm$0.0041 & 0.8074$\pm$0.0084 & 0.1932$\pm$0.0263 \\
SkelCon \cite{tan2022retinal} & 0.9740$\pm$0.0050 & 0.9549$\pm$0.0034 & 0.8352$\pm$0.0170 & 0.9679$\pm$0.0025 & 0.7836$\pm$0.0156 & \underline{0.8335$\pm$0.0102} & \underline{0.1552$\pm$0.0021} \\
SEASA \cite{feng2023unsupervised} & 0.9555$\pm$0.0063 & 0.9444$\pm$0.0029 & 0.8100$\pm$0.0136 & 0.9589$\pm$0.0046 & 0.7401$\pm$0.0075 & 0.8121$\pm$0.0059 & 0.1875$\pm$0.0204 \\
SePiCo \cite{xie2023sepico} & 0.9579$\pm$0.0054 & 0.9380$\pm$0.0020 & 0.8212$\pm$0.0051 & 0.9506$\pm$0.0026 & 0.7213$\pm$0.0055 & 0.8004$\pm$0.0045 & 0.2056$\pm$0.0133 \\
DCLPS \cite{lu2024domain} & \underline{0.9744$\pm$0.0017} & \underline{0.9554$\pm$0.0020} & \underline{0.8361$\pm$0.0192} & \underline{0.9683$\pm$0.0039} & \underline{0.7853$\pm$0.0059} & 0.8320$\pm$0.0080 & 0.1689$\pm$0.0081 \\
\Xhline{1pt}
\textbf{IDA (Ours)} & \textbf{0.9784$\pm$0.0011} & \textbf{0.9595$\pm$0.0010} & \textbf{0.8477$\pm$0.0076} & \textbf{0.9717$\pm$0.0013} & \textbf{0.8038$\pm$0.0042} & \textbf{0.8446$\pm$0.0036} & \textbf{0.1422$\pm$0.0075} \\
\Xhline{1pt}
\end{tabular}}
\end{minipage}

\end{figure*}
\begin{figure*}
    \centering
    \includegraphics[width=12cm]{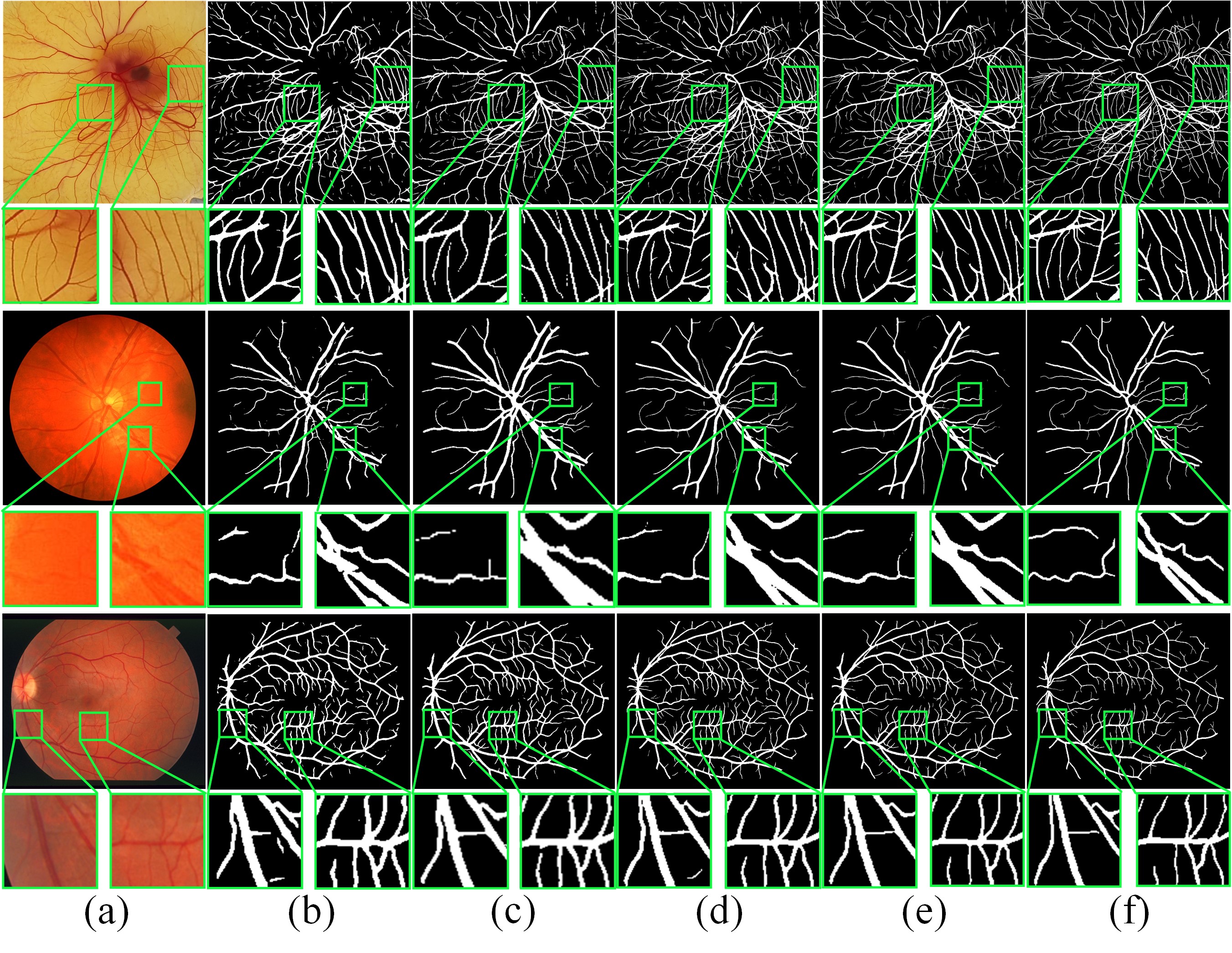}
    \caption{Segmentation results of different methods. (a) Original images. (b) MPSCL \cite{liu2022margin}. (c) SkelCon \cite{tan2022retinal}. (d) DCLPS \cite{lu2024domain}. (e) Ours. (f) Ground truth. From top to bottom: DRIVE → CAM\_DB, DRIVE → CHASEDB1, and DRIVE → STARE. Below each image is an enlarged patch of the image.}\label{figure5}
\end{figure*}

\textbf{\textit{3) Implementation Details:}}
The experimental setup in this paper is as follows: We use W-Net \cite{galdran2022state} for both student and teacher models. The batch size is set to 4. The optimizer is AdamW \cite{loshchilov2017decoupled} with betas $(0.9, 0.999)$ and a weight decay of 5e-4. The learning rate is initially set to 2.5e-4. During training, both the downsampled images and cropped patches from the source and target domains are resized to $384 \times 384$. For ROSSA \cite{ning2024accurate}, since its original image resolution is $320\times320$, we directly use the original resolution as input. During testing, for CAM\_DB, we resize the original images to $512 \times 512$. For CHASEDB1 \cite{fraz2012ensemble} and STARE \cite{hoover2000locating}, the images are resized to $384 \times 384$ before being fed into the student model, and the segmentation outputs are then interpolated back to their original size. For a fair comparison, we adopt a unified implementation framework where all methods are trained with the same backbone and default training settings. Specifically, for CHASEDB1 and STARE, we perform five-fold cross-validation to mitigate biases from data partitioning. For CAM\_DB, where annotations are incomplete, we instead conduct five independent runs with different random seeds to account for randomness. In all cases, we report the mean and standard deviation of the results.

\subsection{Comparison with Existing Methods}
To evaluate the performance of our model, we compare it with several state-of-the-art methods, including adversarial learning-based approaches (AdvEnt \cite{vu2019advent}), contrastive learning- based methods (SkelCon \cite{tan2022retinal}, SePiCo \cite{xie2023sepico}, DCLPS \cite{lu2024domain}), and hybrid strategies that combine both adversarial and contrastive learning (SEASA \cite{feng2023unsupervised}, MPSCL \cite{liu2022margin}).

\textbf{ \textit{1) Results from DRIVE to CAM\_DB:}} Table \ref{Table2} presents the quantitative results for CAM segmentation, where DRIVE serves as the source domain and CAM\_DB as the target domain. We select DRIVE as the source domain due to its greater suitability compared to other retinal datasets, as further evidenced by the ablation study presented later. The best result is highlighted in bold, while the second-best result is underlined. Our method demonstrates superior performance across seven evaluation metrics, achieving a DICE score exceeding 0.89 and scores above 0.93 in the remaining four pixel-level metrics as well as in clDICE. This confirms the feasibility of transferring knowledge from retinal datasets to the CAM dataset. Moreover, our method consistently outperforms competing approaches, particularly in the DICE, clDICE, and BM metrics, with average improvements over the second-best methods of $2.79\%$, $1.88\%$, and $1.61\%$, respectively. These gains demonstrate that our method not only improves the accuracy of vessel pixel classification, but also more effectively preserves vascular continuity and topological integrity, both of which are essential for reliable CAM image segmentation.

\textbf{ \textit{2) Results from DRIVE to CHASEDB1 and STARE:}} To further assess the effectiveness, generalizability, and robustness of our method, we employ the CHASEDB1 and STARE datasets as target domains for cross-domain evaluation. These two datasets present distinct imaging conditions and anatomical characteristics compared to DRIVE, such as differences in resolution, illumination, and vessel distribution. Notably, we do not perform experiments from DRIVE to FIVES or ROSSA, primarily due to computational constraints and the substantial domain gap between ROSSA and DRIVE. As reported in Table \ref{Table3} and Table \ref{Table4}, our method consistently achieves the best overall performance across all evaluated metrics. For pixel-level classification accuracy, it attains the highest scores in AUC, ACC, SE, and SP on both datasets and surpasses the second-best methods by $1.38\%$ and $1.85\%$ in DICE on CHASEDB1 and STARE, respectively, highlighting its superior capability to accurately identify vessel pixels and delineate their spatial boundaries. In addition, our method demonstrates strong performance in preserving topological structures, achieving the best clDICE and BM scores. Specifically, it improves upon the second-best method by $0.99\%$ on CHASEDB1 and $1.11\%$ on STARE in clDICE. These results indicate that our model maintains finer vessel continuity and connectivity under domain shifts, contributing to more reliable anatomical representations.

\setlength{\tabcolsep}{2pt}   
\begin{table*}[1\textwidth]  
\centering
\caption{Ablative Results of Different Source Datasets on the CAM\_DB Task. The best result is marked as \textbf{bold}. ``$\downarrow$'' indicates lower is better.}
\myfont
\fontsize{11pt}{17pt}\selectfont      
\resizebox{1\linewidth}{!}{       
\begin{tabular}{c !{\vrule width 0.7pt} c !{\vrule width 0.7pt} c c c c c c c}
\Xhline{1pt}
\rowcolor[HTML]{C0C0C0} 
Source & Modality & AUC & ACC & SE & SP & DICE & clDICE & BM $\downarrow$ \\
\Xhline{1pt}
DRIVE \cite{staal2004ridge} & Fundus & 0.9854$\pm$0.0010 & \textbf{0.9608$\pm$0.0021} & \textbf{0.9334$\pm$0.0027} & 0.9666$\pm$0.0025 & \textbf{0.8929$\pm$0.0052} & \textbf{0.9388$\pm$0.0015} & \textbf{0.4878$\pm$0.0059} \\
CHASEDB1 \cite{fraz2012ensemble} & Fundus & \textbf{0.9872$\pm$0.0010} & \underline{0.9544$\pm$0.0011} & 0.8535$\pm$0.0178 & \textbf{0.9758$\pm$0.0051} & \underline{0.8677$\pm$0.0006} & \underline{0.9219$\pm$0.0034} & 0.5022$\pm$0.0079 \\
STARE \cite{hoover2000locating} & Fundus & 0.9849$\pm$0.0015 & 0.9520$\pm$0.0051 & \underline{0.8871$\pm$0.0198} & 0.9657$\pm$0.0098 & 0.8663$\pm$0.0105 & 0.9189$\pm$0.0050 & 0.5068$\pm$0.0067 \\
FIVES \cite{jin2022fives} & Fundus & \underline{0.9864$\pm$0.0005} & 0.9528$\pm$0.0012 & 0.8521$\pm$0.0088 & \underline{0.9741$\pm$0.0015} & 0.8633$\pm$0.0039 & 0.9154$\pm$0.0021 & \underline{0.4950$\pm$0.0049} \\
ROSSA \cite{ning2024accurate} & OCTA & 0.9775$\pm$0.0019 & 0.9398$\pm$0.0044 & 0.8627$\pm$0.0233 & 0.9562$\pm$0.0100 & 0.8341$\pm$0.0073 & 0.8744$\pm$0.0054 & 0.5172$\pm$0.0050 \\
\Xhline{1pt}
\end{tabular}}
\label{Table5}
\end{table*}

\textbf{ \textit{3) Visualization:}} Fig. \ref{figure5} presents qualitative comparisons of segmentation results across three domain adaptation tasks: DRIVE → CAM\_DB, DRIVE → CHASEDB1, and DRIVE → STARE. It can be observed that MPSCL\cite{liu2022margin} exhibits both over-segmentation and under-segmentation, resulting in inaccurate delineation of large vessels. SkelCon \cite{tan2022retinal} produces relatively good visual results but tends to generate overconfident predictions, causing vessel regions to appear noticeably thicker than the ground truth. DCLPS \cite{lu2024domain} suffers from disconnections along vascular structures, compromising anatomical continuity. In contrast, our method outperforms others in most cases, including accurate identification of large vessels and vascular branching structures, while preserving overall vessel connectivity. Nevertheless, some limitations remain in challenging regions, such as over-segmentation in CHASEDB1 and under-segmentation in low-contrast regions of STARE, as highlighted in Fig. \ref{figure5}. These cases suggest opportunities for further improvement in handling ambiguous boundaries.

\subsection{Ablation Studies}
In this section, we first identify DRIVE as the optimal source dataset for the CAM\_DB task through comparative experiments. Based on this, we perform ablation studies on the DRIVE → CAM\_DB task to assess the contributions of each component in our IDA framework. Notably, similar conclusions can also be drawn from the DRIVE → CHASEDB1 and DRIVE → STARE tasks.

\textbf{\textit{ 1) Evaluation of Source Datasets:}} To determine the optimal source domain for adaptation to CAM\_DB, we evaluate five retinal image datasets: DRIVE \cite{staal2004ridge}, CHASEDB1 \cite{fraz2012ensemble}, STARE \cite{hoover2000locating}, FIVES \cite{jin2022fives}, and ROSSA \cite{ning2024accurate}. The experimental results are presented in Table \ref{Table5}. DRIVE outperforms others in five of seven metrics, achieving ACC of 0.9608, SE of 0.9334, DICE of 0.8929, clDICE of 0.9388, and BM of 0.4878 (lower is better). We attribute this to its high image quality, consistent vascular structures, and low intra-domain variance. In comparison, the performance of CHASEDB1, STARE, and FIVES is slightly inferior compared to DRIVE, mainly due to larger inter-domain gaps. Specifically, CHASEDB1 and STARE are limited by small sample sizes and considerable variability in vessel morphology. Although FIVES contains 800 samples, its results are impacted by increased complexity arising from diverse pathological conditions. Finally, ROSSA, based on OCTA imaging, presents the largest modality gap with CAM images, yielding the lowest scores. The experimental results in Table \ref{Table4} indicate that DRIVE is better suited as the source domain dataset for the CAM vessel segmentation task compared to the other datasets. Notably, even using ROSSA as the source still yields a DICE above 0.83 and a clDICE over 0.87, highlighting the feasibility of leveraging UDA for knowledge transfer from retinal datasets to CAM images.

\begin{figure}
    \centering
    \includegraphics[width=8.4cm]{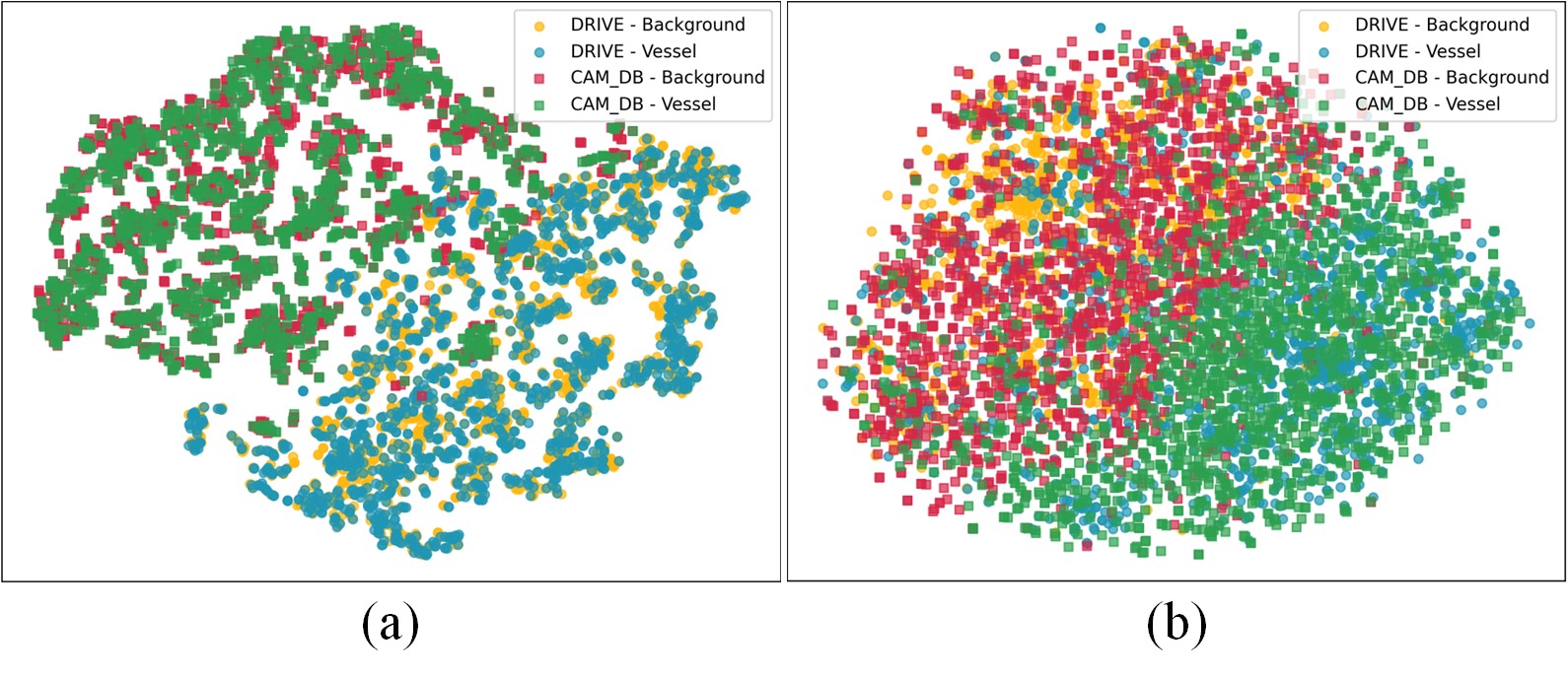}
    \caption{t-SNE visualization of the feature representations on the DRIVE → CAM\_DB task. (a) Significant distribution gap between DRIVE and CAM\_DB before adaptation. (b) Effective feature alignment achieved by IDA, indicating successful domain adaptation.}\label{figure6}
\end{figure}

Furthermore, to better illustrate the impact of domain adaptation, we present t-SNE visualizations of the feature distributions for DRIVE and CAM\_DB before and after adaptation, as shown in Fig. \ref{figure6}. In Fig. \ref{figure6}(a), a clear boundary exists between the two domains, indicating a significant distributional gap. In contrast, Fig. \ref{figure6}(b) shows that, after applying our IDA method, intra-class features are well aligned and inter-class features are distinctly separated, indicating effective domain alignment. These results further confirm that DRIVE is an appropriate choice as the source domain dataset for subsequent ablation studies.

\begin{figure}
    \centering
    \includegraphics[width=8.4cm]{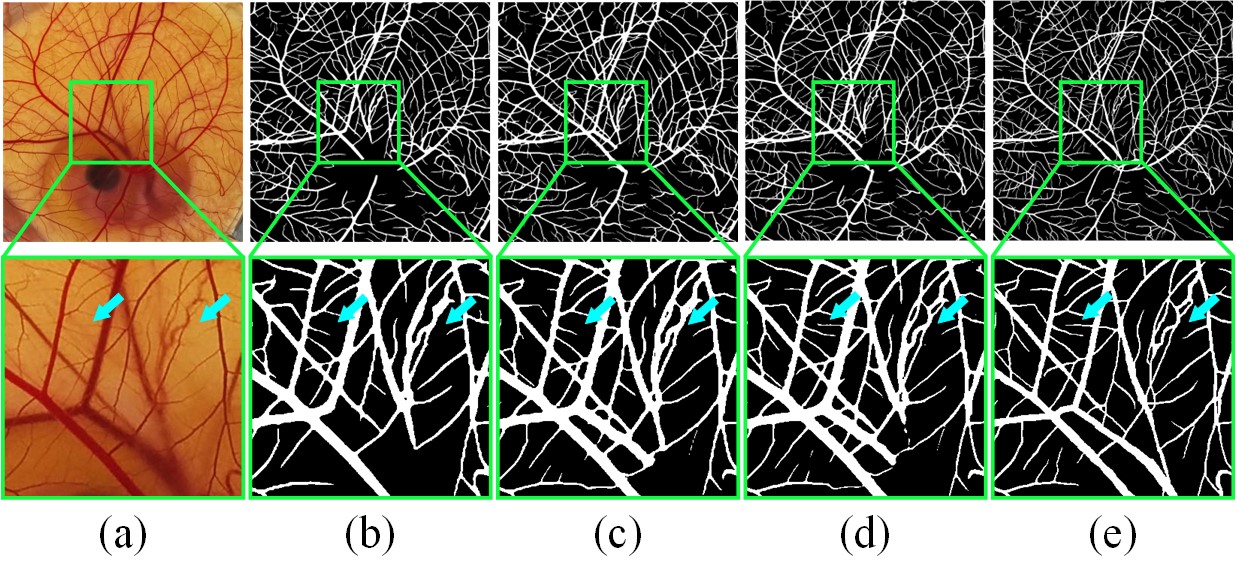}
    \caption{Visualization comparison of different input strategies on the DRIVE → CAM\_DB task. (a) Original images. (b) Patch only. (c) Whole only. (d) Ours. (e) Ground truth. Below each image is an enlarged patch of the image.}\label{figure7}
    \vspace{-10pt}
\end{figure}

\textbf{\textit{ 2) Multi-Resolution Input Strategy:}} “Patch Only” refers to the method that uses random patches from the original image as input, while “Whole Only” uses the scaled full image as input. In our MRAT method, both input strategies are used, denoted as “Patch \& Whole”. As shown in Table \ref{Table6}, the “Patch \& Whole” strategy outperforms the other two in most metrics, indicating strong overall performance in capturing both pixel-level accuracy and structural continuity. These results highlight its superior capability in capturing both pixel-level accuracy and structural continuity. However, the “Patch Only” strategy achieves the best performance in the BM metric, preserving fine-grained topological structures, suggesting that small-scale patch inputs are more effective. In addition, the blue arrows in the visualization results of Fig. \ref{figure7} indicate that our method helps the model capture both the detailed features of vessels and their connectivity, making it more suitable for cross-domain vessel segmentation tasks.

\begin{table}
\centering
\caption{
Ablative results of different input forms in our model on the DRIVE $\rightarrow$ CAM\_DB task. 
}
\myfont
\fontsize{5pt}{8pt}\selectfont
\resizebox{\columnwidth}{!}{
\begin{tabular}{c !{\vrule width 0.15pt} c c c}
\Xhline{0.3pt}
\rowcolor[HTML]{C0C0C0} 
Input Strategy & DICE & clDICE & BM $\downarrow$ \\
\Xhline{0.3pt}
Patch Only & \underline{0.8706$\pm$0.0055} & 0.9136$\pm$0.0053 & \textbf{4.5227$\pm$0.1478} \\
Whole Only & 0.8667$\pm$0.0146 & \underline{0.9357$\pm$0.0035} & 5.0580$\pm$0.1105 \\
Patch \& Whole & \textbf{0.8929$\pm$0.0052} & \textbf{0.9388$\pm$0.0015} & \underline{4.8775$\pm$0.0594} \\
\Xhline{0.3pt}
\end{tabular}}
\par\vspace{2pt}
{\scriptsize\justifying 
\setlength{\parindent}{0pt}
\textbf{Patch Only:} The input comprises randomly extracted patches from the original image. 
\textbf{Whole Only:} The input comprises the entire image, resized to the specified dimensions.\par}
\label{Table6}
\end{table}

\begin{table}[htbp]
\centering
\caption{Ablative Results of Different Training Strategies in Our Model on the DRIVE $\rightarrow$ CAM\_DB Task.}
\myfont
\fontsize{6.5pt}{9pt}\selectfont
\resizebox{\columnwidth}{!}{
\begin{tabular}{c !{\vrule width 0.15pt} c c c}
\Xhline{0.3pt}
\rowcolor[HTML]{C0C0C0} 
Methods \& Settings & DICE & clDICE & BM $\downarrow$ \\
\Xhline{0.3pt}
\multicolumn{4}{c}{Unidirectional Translation (UT)} \\
\Xhline{0.3pt}
S2TCut & 0.8492$\pm$0.0202 & 0.9046$\pm$0.0156 & 0.5147$\pm$0.0074 \\
T2SCut & 0.8449$\pm$0.0104 & 0.9105$\pm$0.0064 & 0.5230$\pm$0.0137 \\
S2TClass & 0.8557$\pm$0.0255 & 0.8996$\pm$0.0187 & 0.5283$\pm$0.0119 \\
T2SClass & 0.8349$\pm$0.0072 & 0.8893$\pm$0.0036 & 0.5388$\pm$0.0101 \\
\Xhline{0.3pt}
\multicolumn{4}{c}{Bidirectional Symmetric Translation (BST)} \\
\Xhline{0.3pt}
Bi-Cut & 0.8587$\pm$0.0070 & \underline{0.9253$\pm$0.0056} & 0.5104$\pm$0.0152 \\
Bi-Class & 0.8625$\pm$0.0121 & 0.9123$\pm$0.0090 & 0.5194$\pm$0.0161 \\
\Xhline{0.3pt}
\multicolumn{4}{c}{Bidirectional Asymmetric Translation (BAT)} \\
\Xhline{0.3pt}
S2TCut \& T2SClass & \underline{0.8746$\pm$0.0144} & 0.9233$\pm$0.0066 & \underline{0.5003$\pm$0.0075} \\
S2TClass \& T2SCut & \textbf{0.8929$\pm$0.0052} & \textbf{0.9388$\pm$0.0015} & \textbf{0.4878$\pm$0.0059} \\
\Xhline{0.3pt}
\end{tabular}}
\par\vspace{2pt}
\scriptsize\justifying 
\setlength{\parindent}{0pt}
\textbf{S2TCut} and \textbf{S2TClass} refer to Unidirectional CutMix and ClassMix from the source domain to the target domain, while \textbf{T2SCut} and \textbf{T2SClass} represent the reverse direction, from the target domain to the source domain. \textbf{Bi-Cut} and \textbf{Bi-Class} represent bidirectional CutMix and ClassMix between the source and target domains, respectively.
\label{Table7}
\end{table}

\textbf{\textit{ 3) Asymmetric Translation Strategy:}} Our MRAT method constructs intermediate images between the source and target domains using CP operations, innovatively employing the Bidirectional Asymmetric Translation (BAT) strategy. Table \ref{Table7} compares BAT with two alternative strategies: Unidirectional Translation (UT) and Bidirectional Symmetric Translation (BST). The UT strategy shows the lowest performance due to its one-way design, which fails to establish consistent learning mechanisms across domains and thus results in semantic mismatches. The BST strategy, while considering both directions, treats source and target domains symmetrically and ignores the fact that target images lack the ground truth. This may introduce ambiguous semantic content into the mixed images, leading to suboptimal performance. In contrast, the BAT strategy achieves superior performance. In particular, the “S2TClass \& T2Scut” setting yields the best overall results. Compared to the second-best setting, it achieves improvements of 1.83\%, 1.55\%, and 1.25\% in DICE, clDICE, and BM, respectively. We attribute this to the higher accuracy of source labels compared to the pseudo-labels, making ClassMix more suitable for source patches. Overall, our method effectively reduces image-level domain shifts and facilitates more robust cross-domain feature learning.

\begin{table}[htbp]
\centering
\caption{Ablative results of each mini-patch size in our Model on the DRIVE $\rightarrow$ CAM\_DB Task.}
\myfont
\fontsize{5.5pt}{8pt}\selectfont
\resizebox{\columnwidth}{!}{
\begin{tabular}{c !{\vrule width 0.15pt} c c c}
\Xhline{0.3pt}
\rowcolor[HTML]{C0C0C0} 
Mini-Patch Size & DICE & clDICE & BM $\downarrow$ \\
\Xhline{0.3pt}
32 & \underline{0.8859$\pm$0.0047} & \underline{0.9273$\pm$0.0046} & 0.5251$\pm$0.0174 \\
64 & 0.8685$\pm$0.0211 & 0.9209$\pm$0.0142 & \underline{0.5149$\pm$0.0191} \\
96 & 0.8611$\pm$0.0175 & 0.9222$\pm$0.0113 & 0.5397$\pm$0.0163 \\
128 & \textbf{0.8929$\pm$0.0052} & \textbf{0.9388$\pm$0.0015} & \textbf{0.4878$\pm$0.0059} \\
192 & 0.8514$\pm$0.0329 & 0.9100$\pm$0.0205 & 0.5367$\pm$0.0272 \\
256 & 0.8256$\pm$0.0458 & 0.8941$\pm$0.0328 & 0.5496$\pm$0.0426 \\
\Xhline{0.3pt}
\end{tabular}}
\label{Table8}
\end{table}

\begin{table}[htbp]
\centering
\caption{Ablative results of different contrastive learning methods in our model on the DRIVE $\rightarrow$ CAM\_DB Task.}
\myfont
\fontsize{4.5pt}{7pt}\selectfont
\resizebox{\columnwidth}{!}{
\begin{tabular}{c !{\vrule width 0.15pt} c c c}
\Xhline{0.3pt}
\rowcolor[HTML]{C0C0C0} 
CL Methods & DICE & clDICE & BM $\downarrow$ \\
\Xhline{0.3pt}
Vanilla CL & 0.8467$\pm$0.0144 & 0.9094$\pm$0.0079 & 0.5143$\pm$0.0055 \\
DCL & \underline{0.8676$\pm$0.0191} & \underline{0.9216$\pm$0.0143} & \underline{0.5120$\pm$0.0047} \\
IDCL (Ours) & \textbf{0.8929$\pm$0.0052} & \textbf{0.9388$\pm$0.0015} & \textbf{0.4878$\pm$0.0059} \\
\Xhline{0.3pt}
\end{tabular}}
\par\vspace{2pt}
\scriptsize\justifying
\setlength{\parindent}{0pt}
\textbf{Vanilla CL:} Prototypes are generated from source features, and both domains perform contrastive learning with source prototypes. \textbf{DCL:} Prototypes are generated by target-like features, and both domains perform contrastive learning with the target-bias prototypes.
\label{Table9}
\end{table}

\begin{table}[htbp]
\centering
\caption{Ablative results of each threshold in our model on the DRIVE $\rightarrow$ CAM\_DB Task.}
\myfont
\fontsize{4pt}{6pt}\selectfont
\resizebox{\columnwidth}{!}{
\begin{tabular}{c c !{\vrule width 0.15pt} c c c}
\Xhline{0.3pt}
\rowcolor[HTML]{C0C0C0} 
$th_{t2s}$ & $th_{s2t}$ & DICE & clDICE & BM $\downarrow$ \\
\Xhline{0.3pt}
0.5 & 0.3 & 0.8620$\pm$0.0355 & 0.9182$\pm$0.0190 & 0.5066$\pm$0.0125 \\
0.5 & 0.5 & 0.8709$\pm$0.0171 & 0.9228$\pm$0.0116 & \underline{0.5028$\pm$0.0088} \\
0.7 & 0.5 & \underline{0.8726$\pm$0.0211} & \underline{0.9238$\pm$0.0133} & 0.5068$\pm$0.0091 \\
0.9 & 0.7 & \textbf{0.8929$\pm$0.0052} & \textbf{0.9388$\pm$0.0015} & \textbf{0.4878$\pm$0.0059} \\
0.9 & 0.9 & 0.8695$\pm$0.0182 & 0.9222$\pm$0.0117 & 0.5097$\pm$0.0115 \\
\Xhline{0.3pt}
\end{tabular}}
\label{Table10}
\end{table}

\textbf{\textit{ 4) Selection of Mini-Patch Size:}} As shown in Table \ref{Table8}, we evaluate the impact of different values of the mask sizes, $m$, for $M_R$ and $M_C$ in the MRAT strategy on the performance for the DRIVE → CAM\_DB task. It can be observed that when the mask size $m$ is set to 128, all metrics achieve optimal performance, with the most significant improvements observed in the clDICE and BM metrics. This indicates that the model is better able to integrate semantic information from the source and target domains, thus achieving image-level adaptation.

\textbf{\textit{5) Intermediate Domain-guided Contrastive Learning Module:}} To demonstrate the effectiveness of our IDCL with traditional contrastive learning (where prototypes are generated from source features, and both domains perform contrastive learning with source prototypes, referred to as Vanilla CL) and the domain-interactive contrastive learning (DCL) method in DCLPS \cite{lu2024domain}. For a fair comparison, all contrastive learning methods are implemented within the IDA framework proposed in this paper. As shown in Table \ref{Table9}, Vanilla CL performs poorly, while our IDCL achieves the best performance. This is because Vanilla CL primarily focuses on enhancing intra-domain class separability but neglects inter-domain class discrepancies, limiting the model’s ability to effectively learn features from the target domain. Although DCL accounts for domain shifts, it directly uses target features for contrastive learning, overlooking the need for consistency between prototypes and features from both domains in the embedding space. In contrast, our IDCL fully utilizes intermediate domain information and establishes a consistent learning strategy for both the source and target domains. This enables the smooth interaction of cross-domain information, resulting in a notable improvement in performance.

\textbf{\textit{ 6) Selection of Thresholds:}} Considering the uncertainty of pseudo-labels in the target domain, we employ dynamically adjusted prototype weights, $w_{t2s}^{(k)}$ and $w_{s2t}^{(k)}$, in our IDCL. Table \ref{Table10} presents the impact of different threshold hyperparameter settings in Eq. \ref{equation10} on segmentation performance. Here, $th_{t2s}$ denotes the case where patches from the target domain are mixed into source images to form the intermediate images, while $th_{s2t}$ represents the reverse. Since we employ a pretrained model, the segmentation confidence on source domain images is initially higher than that on target domain images images during the early stages of domain adaptation. Consequently, the best performance across all three metrics is achieved when $th_{t2s} = 0.9$ and $th_{s2t} = 0.7$

\begin{table}[htbp]
\centering
\caption{Ablative results of different prototype update strategies in our model on the DRIVE $\rightarrow$ CAM\_DB Task.}
\myfont
\fontsize{6pt}{8.5pt}\selectfont
\resizebox{\columnwidth}{!}{
\begin{tabular}{c !{\vrule width 0.15pt} c c c}
\Xhline{0.3pt}
\rowcolor[HTML]{C0C0C0} 
Prototype Weights & DICE & clDICE & BM $\downarrow$ \\
\Xhline{0.3pt}
0.1 & 0.8292$\pm$0.0187 & 0.8978$\pm$0.0096 & 0.5286$\pm$0.0121 \\
0.3 & 0.8386$\pm$0.0230 & 0.9060$\pm$0.0131 & \underline{0.5158$\pm$0.0120} \\
0.5 & 0.8483$\pm$0.0200 & 0.9095$\pm$0.0115 & 0.5271$\pm$0.0087 \\
0.7 & \underline{0.8533$\pm$0.0213} & \underline{0.9133$\pm$0.0145} & 0.5164$\pm$0.0082 \\
0.9 & 0.8333$\pm$0.0266 & 0.9003$\pm$0.0141 & 0.5283$\pm$0.0165 \\
Dynamic & \textbf{0.8929$\pm$0.0052} & \textbf{0.9388$\pm$0.0015} & \textbf{0.4878$\pm$0.0059} \\
\Xhline{0.3pt}
\end{tabular}}
\par\vspace{2pt}
\label{Table11}
\end{table}

\textbf{\textit{ 7) Effectiveness of Prototype Weights:}} To evaluate the effectiveness of dynamically updating prototype weights, we compare our method with several fixed-weight configurations, as presented in Table \ref{Table11}. When the weights are kept too low throughout the training process, it negatively impacts the model’s learning. This is because the prototypes are unable to absorb new knowledge in each iteration, resulting in an under-represented feature. Conversely, when the weights are kept too high, such as when Prototype Weights = 0.9, the presence of false positive pseudo-label errors can introduce considerable noise into the prototypes, thereby reducing the model performance. Our approach, by considering the predicted probabilities of intermediate images at each iteration, mitigates the impact of incorrect pseudo-labels. Furthermore, it enables the prototypes to fully absorb the intermediate features, leading to improved performance.

\begin{table}[htbp]
\centering
\caption{Ablative results of contrastive loss weights in our model on the DRIVE $\rightarrow$ CAM\_DB Task.}
\myfont
\fontsize{3.5pt}{5.5pt}\selectfont
\resizebox{\columnwidth}{!}{
\begin{tabular}{c c !{\vrule width 0.15pt} c c c}
\Xhline{0.3pt}
\rowcolor[HTML]{C0C0C0} 
$\beta_1$ & $\beta_2$ & DICE & clDICE & BM $\downarrow$ \\
\Xhline{0.3pt}
0.5 & 0.1 & 0.8783$\pm$0.0183 & 0.9260$\pm$0.0103 & 0.5012$\pm$0.0068 \\
1.0 & 0.5 & 0.8848$\pm$0.0225 & 0.9284$\pm$0.0147 & 0.4974$\pm$0.0040 \\
1.0 & 1.0 & \textbf{0.8929$\pm$0.0052} & \textbf{0.9388$\pm$0.0015} & \textbf{0.4878$\pm$0.0059} \\
1.5 & 1.0 & \underline{0.8850$\pm$0.0058} & \underline{0.9315$\pm$0.0052} & \underline{0.4894$\pm$0.0097} \\
\Xhline{0.3pt}
\end{tabular}}
\par\vspace{2pt}
\label{Table12}
\end{table}

\begin{table}[htbp]
\centering
\caption{Ablative results of cnsistency loss weights in our model on the DRIVE $\rightarrow$ CAM\_DB Task.}
\myfont
\fontsize{2.5pt}{4pt}\selectfont
\resizebox{\columnwidth}{!}{
\begin{tabular}{c !{\vrule width 0.15pt} c c c}
\Xhline{0.3pt}
\rowcolor[HTML]{C0C0C0} 
$\gamma$ & DICE & clDICE & BM $\downarrow$ \\
\Xhline{0.3pt}
0.1 & 0.8628$\pm$0.0192 & 0.9208$\pm$0.0095 & \underline{0.5028$\pm$0.0043} \\
0.5 & \underline{0.8713$\pm$0.0201} & \underline{0.9259$\pm$0.0123} & 0.5040$\pm$0.0082 \\
1.0 & \textbf{0.8929$\pm$0.0052} & \textbf{0.9388$\pm$0.0015} & \textbf{0.4878$\pm$0.0059} \\
1.5 & 0.8629$\pm$0.0260 & 0.9203$\pm$0.0145 & 0.5081$\pm$0.0045 \\
\Xhline{0.3pt}
\end{tabular}}
\label{Table13}
\end{table}

\textbf{\textit{ 8) Selection of Loss Function Weights:}}  To determine the optimal combination of loss function weights, we conduct ablation studies on the contrastive loss weights and the consistency loss weight defined in Eq. \ref{equation16}, with the results summarized in Table \ref{Table12} and Table \ref{Table13}, respectively. Notably, when analyzing the effect of contrastive loss weights, the consistency loss weight $\gamma$ is fixed to 1. Conversely, when evaluating the effect of consistency loss weights, the contrastive loss weights $\beta_1$ and $\beta_2$ are fixed to 1. The results indicate that setting all loss weights to 1 yields the best overall performance across all metrics, further demonstrating the robustness of our method and its effectiveness without requiring extensive hyperparameter tuning.

\begin{table}[htbp]
\centering
\caption{Ablative Results of Each Pre-training Strategy in Our Model on the DRIVE $\rightarrow$ CAM\_DB Task.}
\myfont
\fontsize{5.5pt}{9pt}\selectfont
\resizebox{\columnwidth}{!}{
\begin{tabular}{c c c !{\vrule width 0.15pt} c c c}
\Xhline{0.3pt}
\rowcolor[HTML]{C0C0C0} 
Random & Self-Cut & VCL & DICE & clDICE & BM $\downarrow$ \\
\Xhline{0.3pt}
\checkmark &         &         & 0.8374$\pm$0.0101 & 0.8988$\pm$0.0078 & 0.5367$\pm$0.0095 \\
\checkmark & \checkmark &         & \underline{0.8696$\pm$0.0167} & \underline{0.9205$\pm$0.0074} & \underline{0.5123$\pm$0.0071} \\
\checkmark &         & \checkmark & 0.8565$\pm$0.0052 & 0.9128$\pm$0.0045 & 0.5263$\pm$0.0031 \\
\checkmark & \checkmark & \checkmark & \textbf{0.8929$\pm$0.0052} & \textbf{0.9388$\pm$0.0015} & \textbf{0.4878$\pm$0.0059} \\
\Xhline{0.3pt}
\end{tabular}}
\label{Table14}
\end{table}

\textbf{\textit{ 9) Choice of Pre-training Strategies: }} In our default settings, both the student and teacher models are initialized using pre-trained models from the source dataset, with the teacher model’s parameters continuously updated through the EMA strategy. We investigate the impact of different pre-training strategies on the final performance, as shown in Table \ref{Table14}. The “Random” strategy initializes model weights randomly and then trains the model on the source dataset to obtain the pre-trained model. The “Self-Cut” strategy involves applying the multi-resolution CutMix operation exclusively to source images, which can be expressed by the formula $x_{sp} \odot M_R + x_{sr} \odot (\mathbf{1} - M_R)$. The "VCL" strategy implements Vanilla CL within the source domain. Compared to other pre-training strategies, the combination of the "Self-Cut" with the "VCL" proves more effective in enhancing the model performance.

\begin{table}[htbp]
\centering
\caption{Ablative results of each component in our model on the DRIVE $\rightarrow$ CAM\_DB Task.}
\myfont
\fontsize{3pt}{5pt}\selectfont
\resizebox{\columnwidth}{!}{
\begin{tabular}{c c c c !{\vrule width 0.15pt} c c c}
\Xhline{0.3pt}
\rowcolor[HTML]{C0C0C0} 
$\dagger$ & $\ddagger$ & $\dashv$ & $\vdash$ & DICE & clDICE & BM $\downarrow$ \\
\Xhline{0.3pt}
         &         &         &         & 0.7558$\pm$0.0080 & 0.8663$\pm$0.0071 & 0.5713$\pm$0.0101 \\
\checkmark &         &         &         & 0.7610$\pm$0.0225 & 0.8714$\pm$0.0115 & 0.5562$\pm$0.0200 \\
\checkmark & \checkmark &         &         & 0.8134$\pm$0.0150 & 0.8953$\pm$0.0075 & 0.5403$\pm$0.0120 \\
\checkmark & \checkmark & \checkmark &         & \underline{0.8853$\pm$0.0164} & \underline{0.9276$\pm$0.0115} & \underline{0.4967$\pm$0.0070} \\
\checkmark & \checkmark & \checkmark & \checkmark & \textbf{0.8929$\pm$0.0052} & \textbf{0.9388$\pm$0.0015} & \textbf{0.4878$\pm$0.0059} \\
\Xhline{0.3pt}
\end{tabular}}
\par\vspace{2pt}
\scriptsize\justifying
\setlength{\parindent}{0pt}
“$\dag$” refers to a self-training strategy, where a teacher model generates pseudo labels for the target domain, and a student model is supervised using source ground truth and target pseudo labels. “$\ddag$” means our MRAT strategy, “$\dashv$” denotes our IDCL module, and “$\vdash$” represents the aforementioned pre-training strategy.
\label{Table15}
\end{table}

\textbf{\textit{ 10) Ablation of Different Components: }} To validate the effectiveness of each component in our method, we conduct ablation experiments, as shown in Table \ref{Table15}. The baseline in the first row, trained solely on the source domain and evaluated on the target domain, yields suboptimal performance due to the domain shift. By incorporating a simple self-training strategy, we observe a measurable improvement, which is further supported by the reduction in the BM metric, which reflects better preservation of the target domain’s topological structure. Moreover, the integration of our proposed modules (MRAT and IDCL) consistently and substantially improves performance across all evaluation metrics by effectively reducing domain gaps at both the image and feature levels. Furthermore, the utilization of our pretraining strategy enhances the quality of pseudo labels, thereby enabling the model to achieve optimal overall performance.

\section{Conclusion}
In this paper, we propose a novel unsupervised CAM vessel segmentation framework, IDA, which utilizes labeled retinal datasets to assist in the automatic and accurate segmentation of CAM vessels. Our IDA method generates intermediate images through the MRAT strategy to facilitate image-level interaction. To fully leverage the intermediate information, we design a consistent learning strategy for both the source and target domains, using intermediate features to update prototypes that guide domain-wise contrastive learning. This enables intra- class compactness and inter-class separability within domains and allows for feature-level adaptation through cross-domain feature interaction. Comprehensive experiments on the DRIVE → CAM\_DB task demonstrate that our method surpasses existing approaches, achieving superior performance with enhancements of 2.79\%, 1.88\%, and 1.61\% over the second-best results in DICE, clDICE, and BM metrics, respectively. Furthermore, our approach consistently attains the highest scores across all seven evaluation metrics on the DRIVE → CHASEDB1 and DRIVE → STARE tasks, highlighting its strong generalizability and robustness. In the future, we plan to apply this model to angiogenesis-related research for rapid and accurate quantitative parameter characterization, while laying a foundation for future investigations into the effects of various experimental factors on embryonic vascular development in CAM models.

\section*{Acknowledgement}
This work was supported by the National Key R\&D Program of China (Grant No. 2022YFA1104800) and the Anhui Provincial Natural Science Foundation (Grant No. 2308085MF219)
  













\end{document}